\documentclass[aps,prd,twocolumn,nofootinbib,showpacs,preprintnumbers,floatfix]{revtex4}
\usepackage{bm,amsmath,amssymb}
\usepackage{graphicx}
\usepackage{subfigure}
\usepackage{color}
\usepackage{hyperref}
\usepackage{multirow}
\usepackage{lineno}
\usepackage{url}

\newcommand \ber {\begin{eqnarray}}
\newcommand \eer {\end{eqnarray}}
\newcommand \beq {\begin{equation}}
\newcommand \eeq {\end{equation}}
\newcommand \nn  {\nonumber}
\newcommand \sgh {\sigma(\omega,\vec{p},T)}

\begin{document}

\title{In-medium modifications of open and hidden strange-charm mesons\\
from spatial correlation functions}

\author{A. Bazavov}
\affiliation{Department of Physics and Astronomy, University of Iowa, Iowa City, IA
52240, USA}
\author{F. Karsch}
\affiliation{Fakult\"at f\"ur Physik, Universit\"at Bielefeld, D-33615 Bielefeld,
Germany}
\affiliation{Physics Department, Brookhaven National Laboratory, Upton, NY 11973, USA}
\author{Y. Maezawa}
\affiliation{Fakult\"at f\"ur Physik, Universit\"at Bielefeld, D-33615 Bielefeld,
Germany}
\author{Swagato Mukherjee}
\affiliation{Physics Department, Brookhaven National Laboratory, Upton, NY 11973, USA}
\author{P. Petreczky}
\affiliation{Physics Department, Brookhaven National Laboratory, Upton, NY 11973, USA}

\begin{abstract}
We calculate spatial correlation functions of in-medium mesons consisting of
strange--anti-strange, strange--anti-charm and charm--anti-charm quarks in
(2+1)-flavor lattice QCD using the highly improved staggered quark action. 
A comparative study of the in-medium modifications 
of mesons with different flavor contents is performed. We observe
significant in-medium modifications for the $\phi$ and $D_s$ meson channels already
at temperatures around the chiral crossover region. On the other hand, for the
$J/\psi$ and $\eta_c$ meson channels in-medium modifications remain relatively small
around the chiral crossover region and become significant only above 1.3 times the
chiral crossover temperature.

\end{abstract}

\date{\today}
\maketitle

\section{Introduction}

At high temperatures matter controlled by the strong force undergoes a chiral
crossover transition \cite{Bhattacharya:2014ara}, accompanied by the deconfinement of
flavor quantum numbers carrying degrees of freedom \cite{Bazavov:2013dta}.  The
relevant degrees of freedom change from hadrons to quarks and gluons (see {\sl e.g.}
Refs.~\cite{Petreczky:2012rq,Philipsen:2012nu} for recent reviews).  The in-medium
modification and dissolution of heavy quarkonium were suggested as a signal for
creating a deconfined medium in heavy ion collisions by Matsui and Satz
\cite{Matsui:1986dk}.  The existence of heavy-light mesons above the chiral
transition temperature has also been proposed to explain the large energy loss and
flow of heavy quarks observed in heavy ion collisions \cite{Sharma:2009hn}.  Recent
lattice QCD calculations suggest that heavy-light bound states
dissolve already at or close to the QCD transition temperature
based on flavor and quantum number correlation analysis
\cite{Bazavov:2014yba}. 

Hadronic correlation functions have long been
advocated as convenient tools to explore the properties of strong interaction
matter \cite{DeTar:1987ar,DeTar:1987xb}. They encode the in-medium properties of
hadrons, as well as their dissolution.  Moreover, through the comparison of lattice
results with weak coupling calculations at high temperature
\cite{Laine:2003bd,Brandt:2014uda} they also provide information on the change from
strongly to weakly interacting matter. 

Spectral functions, the Fourier transforms of real time meson correlation function,
are the basic quantities that provide knowledge regarding the in-medium properties of mesons
and their dissolution. Meson states appear as peaks in the corresponding spectral
functions with the peak position equal to the meson mass.  The width of the peak
corresponds to the in-medium width of the meson.  However, lattice QCD is formulated
in Euclidean space time. Temporal meson correlation functions calculated on the
lattice, 
\beq
G(\tau, \vec{p},T) = \int d^3x e^{i \vec{p}\cdot\vec{x}} 
\langle J_H(\tau, \vec{x}) J_H(0,
\vec{0}) \rangle \; ,
\eeq
have a simple relation to the spectral function, $\sigma(\omega, \vec{p},T)$:
\ber
G(\tau, \vec{p},T) &=& \int_0^{\infty} d \omega
\sgh K(\omega, \tau, T), \label{eq.spect} \nn\\
K(\omega, \tau, T) &=& \frac{\cosh(\omega(\tau-1/2
T))}{\sinh(\omega/2 T)}.
\label{eq.kernel}
\eer
Here $J_H$ is a meson operator, typically of the form $J_H=\bar q \Gamma_H q$, with
$\Gamma_H$ being some combination of the Dirac matrices that specifies the quantum
numbers of the meson. One way to obtain the spectral function from the above relation
is to use the maximum entropy method 
\cite{Asakawa:2000tr,Wetzorke:2001dk,Asakawa:2002xj,Asakawa:2003re,Datta:2003ww,Jakovac:2006sf,Aarts:2007pk,Ding:2012sp}.
The analysis of temporal correlation functions is difficult due to the limited
extent, $1/T$, in Euclidean time direction.  In the case of heavy quarkonium
correlators, for instance, it turned out that the melting of bound states does not
lead to large changes in the correlation functions
\cite{Petreczky:2008px,Ding:2012sp}. In order to become sensitive to the
corresponding disappearance of a resonance peak in the spectral function high
statistical accuracy and the analysis of the correlation function at a large number
of Euclidean time separations are needed. At fixed temperature $T=1/N_\tau a$, this
requires large lattices with temporal extent $N_\tau$ and sufficiently small lattice
spacing, $a$.

Alternatively, one can study the spatial correlation functions of mesons 
\beq
G(z,T) = \int_0^{1/T} d \tau \int dx dy
\langle J_H(\tau,x,y,z) J_H(0,0,0,0) \rangle.
\eeq
These are related  to the spectral functions in a more complicated way that also
involves integration of momenta,
\beq
G(z,T)=\int_{0}^{\infty} \frac{2 d \omega}{\omega} \int_{-\infty}^{\infty} d p_z e^{i p_z z} \sigma(\omega,p_z,T).
\label{eq:spatial}
\eeq
Since the spatial separation is not limited by the inverse temperature, the spatial
correlation function can be studied at separations larger than $1/T$.  Therefore, the
spatial correlation functions can be more sensitive to in-medium modifications and/or
the dissolution of mesons. Another advantage of spatial correlation functions over
the temporal ones is that the spatial correlation function can be directly compared
to the corresponding vacuum correlation function to quantify modifications of the
in-medium spectral function. It is apparent from Eq.~(\ref{eq.kernel}) that for the
temporal correlation function the temperature dependent in-medium modifications of
the spectral function are partly masked by the temperature dependence of the Kernel
$K(\omega,\tau,T)$, and a comparison with the corresponding vacuum correlation
function demands evaluation of the reconstructed correlator \cite{Datta:2003ww}. Such
complication is not present for the spatial correlation functions. 

While the general relation between spectral functions and spatial meson correlators
is more involved, in some limiting cases it becomes simple. At large distances the
spatial correlation functions decay exponentially, $G(z,T) \sim \exp(-M(T)z)$, where
$M$ is known as the \emph{screening mass}. Note that, unlike the in-medium temporal
correlation functions, the transport contributions to the spectral functions at small
frequencies do not lead to a non-decaying constant in the large distance behavior of
the spatial correlation functions.  At small enough temperatures when there exists a
well-defined mesonic bound state, the spectral function has the form
$\sigma(\omega,0,0,p_z,T)\sim\delta(\omega^2-p_z^2-m_0^2)$, and $M$ becomes equal to
the (pole) mass $m_0$ of the meson. On the other hand, at high enough temperatures,
when the mesonic excitations are completely melted, the spatial meson correlation
functions describe the propagation of a free quark-antiquark pair. The screening
masses are then given by \cite{Florkowski:1993bq}
\begin{equation}
M_{\rm free}=\sqrt{m_{q_1}^2+(\pi T)^2}+ \sqrt{m_{q_2}^2+(\pi T)^2} \; ,
\label{eq:Mfree}
\end{equation}
where $m_{q_1}$ and $m_{q_2}$ are the masses of the quark and anti-quark that form
the meson. This form of the screening mass in the non-interacting limit is a direct
consequence of the anti-periodic boundary conditions in Euclidean time that are
needed for the representation of fermions at non-zero temperature. This leads to the
appearance of a smallest non-zero Matsubara frequency, $\pi T$, in the quark and
anti-quark propagators.  As the bosonic meson state is dissolved in the
non-interacting limit the screening mass results as the contribution of two
independently propagating fermionic degrees of freedom.  Thus the transition between
these two limiting values of the screening mass can be used as an indicator for the
thermal modification and eventual dissolution of mesonic excitations.

Lattice QCD studies of the screening masses of light quark mesons have been performed
within the quenched approximation \cite{Kaczmarek:2013kva,Gavai:2008yv} and also with
two dynamical flavors using staggered \cite{Banerjee:2011yd} as well as Wilson-type
quarks \cite{Iida:2010jz}. Screening masses in the light and strange quark sector
have been studied recently in ($2+1$)-flavor QCD using the so-called p4 staggered
fermion action \cite{Cheng:2010fe,Laermann:2012sr} and the expected qualitative
behavior discussed above was observed.  Furthermore, the study has been extended to
the case of charmonium providing the first direct evidence for melting of the
charmonium ground state \cite{Karsch:2012na} from lattice QCD with light dynamical
quark degrees of freedom.

In this work we report, for the first time, on studies of spatial meson correlators
and screening masses using the Highly Improved Staggered Quark (HISQ) action
\cite{Follana:2006rc} with a strange quark mass tuned to its physical value and
almost physical, degenerate up and down quark masses. The HISQ action is known to
lead to discretization effects that are smaller than those observed with all other
staggered-type actions currently used in studies of lattice QCD thermodynamics
\cite{Bazavov:2011nk}. Moreover, the HISQ action is well suited to study heavy quarks
on the lattice \cite{Follana:2006rc} and turned out to be successful in quantitative
studies of charmonium \cite{Davies:2013ju} and $D$ meson properties
\cite{Bazavov:2013nfa}.  In this work we study the spatial correlation functions of
mesonic excitations with strange ($s$) and charm ($c$) quarks, specifically the
lowest states in the pseudo-scalar, vector, scalar and axial-vector channels for the
$s\bar{s}$, $s\bar{c}$ and $c\bar{c}$ flavor combinations. 
In the following we refer to the meson states corresponding to these flavor combinations
as hidden strange, open charm-strange and hidden charm mesons.
Some preliminary results from these study
have been published in conference proceedings \cite{Maezawa:2013nxa,Bazavov:2014uda}.

\section{Lattice setup}

We calculate meson correlation functions on gauge configurations generated in
$(2+1)$-flavor QCD using the HISQ action \cite{Bazavov:2011nk}. The strange quark
mass $m_s$ is adjusted to its physical value and the light quark masses are fixed at
$m_l=m_s/20$, corresponding to $m_\pi \simeq 160$ MeV and $m_K \simeq 504$ MeV at
zero temperature in the continuum limit \cite{Bazavov:2011nk}.  Charm quarks are
introduced as valance quarks and we use the HISQ action with the so-called
$\epsilon$-term for the charm quark mass \cite{Follana:2006rc} which makes our
calculations in the heavy quark sector free of tree-level discretization errors up to
${\cal O}((am_c)^4)$.  Our calculations have been performed on lattices of size
$N_\sigma^3\times N_\tau=48^3\times 12$.  We consider lattice couplings in the range
$\beta=$6.664--7.280 which correspond to temperatures $T=$138--248 MeV.  This enables
investigation of in-medium modifications of meson properties below and above the
chiral crossover transition at $T_c=(154\pm9)$ MeV \cite{Bazavov:2011nk}.  To study
the spatial correlators at higher temperatures we adopt the fixed-scale approach and
perform calculations at $\beta=7.280$ for $N_\tau = 10, 8, 6, 4$ which corresponds to
the temperature range $T=$298--744 MeV.  In all our calculations the spatial extent
of the lattice is four times the temporal extent: $N_\sigma=4N_\tau$.  The lattice
spacing and the resulting temperature values, $T=1/N_\tau a$, have been determined
using results for the kaon decay constant \cite{Bazavov:2011nk}. These temperatures
together with the run parameters for all our finite temperature calculations are
summarized in Tab.~\ref{tab:finite}.

\begin{table}[t]
\begin{center}
\caption{Gauge coupling ($\beta$), strange ($m_s$) and charm ($m_c$) quark masses,
temporal lattice sizes ($N_\tau$) and the number of trajectories (traj.) used for the
finite temperature calculations.  The light quark mass is fixed as $m_l=m_s/20$.  The
meson correlation functions are calculated every 10 trajectories.  The spatial
lattice extent is $N_\sigma=4 N_{\tau}$.  We also show the temperature values
determined using $f_K$ as an input.}
\label{tab:finite}
\begin{tabular}{cccccc}
\hline\hline
$\beta$ & $am_s$ & $am_c$ & $N_\tau$ & traj. & $T$ [MeV]  \\
\hline
 6.664 & 0.0514 & 0.632 & 12  & 3740  & 138.2\\
 6.700 & 0.0496 & 0.604 & 12  & 6500  & 143.3\\
 6.740 & 0.0476 & 0.575 & 12  & 6170  & 149.0\\
 6.770 & 0.0460 & 0.554 & 12  & 6320  & 153.5\\
 6.800 & 0.0448 & 0.534 & 12  & 6590  & 158.0\\
 6.840 & 0.0430 & 0.509 & 12  & 7910  & 164.3\\
 6.860 & 0.0420 & 0.497 & 12  & 3660  & 167.5\\
 6.880 & 0.0412 & 0.486 & 12  & 9620  & 170.8\\
 6.910 & 0.0400 & 0.469 & 12  & 4130  & 175.8\\
 6.950 & 0.0386 & 0.448 & 12  & 6200  & 182.6\\
 6.990 & 0.0370 & 0.429 & 12  & 5100  & 189.6\\
 7.030 & 0.0356 & 0.410 & 12  & 6700  & 196.9\\
 7.100 & 0.0332 & 0.380 & 12  & 10050 & 210.2\\
 7.150 & 0.0320 & 0.360 & 12  & 9590  & 220.2\\
 7.280 & 0.0284 & 0.315 & 12  & 11120 & 247.9\\
 7.280 & 0.0284 & 0.315 & 10  & 4180  & 297.5\\
 7.280 & 0.0284 & 0.315 &  8  & 4990  & 371.9\\
 7.280 & 0.0284 & 0.315 &  6  & 3810  & 495.8\\
 7.280 & 0.0284 & 0.315 &  4  & 4820  & 743.7\\
\hline\hline
\end{tabular}
\end{center}
\end{table}

In the staggered formulation  quarks come in four valence tastes and meson operators
are defined as $J_H = \bar{q} ( \Gamma^D \times \Gamma^F ) q$, $\Gamma^D$ and
$\Gamma^F$ being products of Dirac Gamma matrices which generate spin and taste
structures, respectively \cite{Altmeyer:1992dd}.  In this study we focus only on
local meson operators with $\Gamma^D = \Gamma^F = \Gamma$.  By using staggered quark
fields $\chi({\bm x})$ at ${\bm x}=(\tau, x,y,z)$ the local meson operators can be
written in a simple form $J_H({\bm x})=\tilde{\phi}({\bm x}) \bar{\chi}({\bm x})
\chi({\bm x})$, where $\tilde{\phi}({\bm x})$ is a phase factor depending on the
choice of $\Gamma$.  We calculate only the quark-line connected part of the meson
correlators since the contribution of the disconnected part either vanishes or is
expected to be small in most cases considered in this study (see discussions below).
Since a staggered meson correlator couples to two different meson excitations with
opposite parity, the large distance behavior of the lattice correlator can be
described by 
\begin{eqnarray}
G(\tau) = A_{NO}^2 \left( e^{-M_-\tau} + e^{-M_-(N_\tau - \tau)} \right)
\nonumber \\
-(-1)^\tau A_{O}^2 \left( e^{-M_+\tau} + e^{-M_+(N_\tau - \tau)} \right) ,
\label{eq:fit}
\end{eqnarray}
where the first (second) term on the right-hand-side characterizes a non-oscillating
(oscillating) contribution governed by a negative (positive) parity state.  Taking
the square of the amplitudes ensures their positivity \cite{Lepage:2001ym}.  In
Tab.~\ref{tab:status} we summarize the different choices of the phase factor $\tilde
\phi$ and the meson states they correspond to.  We have considered four channels
which we denote as scalar (S), pseudo-scalar (PS), axial-vector (AV) and vector (V).
Notice that the oscillating state does not exist for PS channel of $s\bar{s}$ and
$c\bar{c}$ sectors \cite{Altmeyer:1992dd}, thus we impose $A_O=0$ on these
correlators.  The negative parity states in these channels correspond to different
tastes of the same physical meson and will thus have nearly degenerate masses if
lattice spacings is sufficiently small. For instance, in the $c \bar c$ sector the
negative parity states in S and PS channels both correspond to the same $\eta_c$
state. We will comment on this in more detail later.

\begin{table}[t]
\caption{List of meson operators and corresponding physical states in the strange
($s\bar{s}$), strange-charm ($s\bar{c}$) and charm ($c\bar{c}$) sectors.  The
lightest $s\bar{s}$ pseudo-scalar state is defined as $M_{\eta_{s\bar{s}}} = \sqrt{ 2
M_K^2 - M_\pi^2} \sim686$ MeV which is used to determine the strange quark mass on
the zero-temperature lattices.}
\label{tab:status}
\begin{center}
\begin{tabular}{ccccccc}
\hline\hline
                 &  $- \tilde \phi(x)$                       & $\Gamma$            & $J^{PC}$  & $s\bar{s}$   & $s\bar{c}$    & $c\bar{c}$  \\\hline
$M_{-}^{\rm S}$  & \multirow{2}{*}{$1$}                      & $\gamma_4 \gamma_5$ & $0^{-+}$  & $\eta_{s\bar{s}}$ & $D_s$    & $\eta_c$    \\
$M_{+}^{\rm S}$  &                                           & $ 1 $               & $0^{++}$  &              & $D_{s0}^\ast$ & $\chi_{c0}$ \\\hline
$M_{-}^{\rm PS}$ & \multirow{2}{*}{$(-1)^{x+y+z}$}           & $\gamma_5$          & $0^{-+}$  & $\eta_{s\bar{s}}$ & $D_s$    & $\eta_c$    \\
$M_{+}^{\rm PS}$ &                                           & $\gamma_4$          & $0^{+-}$  & --           &               & --          \\\hline
$M_{-}^{\rm AV}$ & \multirow{2}{*}{$(-1)^{x},~(-1)^{y}$}     & $\gamma_i\gamma_4$  & $1^{--}$  & $\phi$       & $D_s^\ast$    & $J/\psi$    \\
$M_{+}^{\rm AV}$ &                                           & $\gamma_i\gamma_5$  & $1^{++}$  & $f_1(1420)$  & $D_{s1}$      & $\chi_{c1}$ \\\hline
$M_{-}^{\rm V}$  & \multirow{2}{*}{$(-1)^{x+z},~(-1)^{y+z}$} & $\gamma_i$          & $1^{--}$  & $\phi$       & $D_s^\ast$    & $J/\psi$    \\
$M_{+}^{\rm V}$  &                                           & $\gamma_j\gamma_k$  & $1^{+-}$  &              &               & $h_c$       \\
\hline\hline
\end{tabular}
\end{center}
\end{table}

In Eq.~(\ref{eq:fit}) as well as in Tab.~\ref{tab:status} we assumed that the
direction of propagation is the imaginary time $\tau$. When discussing spatial
correlation functions we assume that the direction of propagation is $z$. In that
case, $z$ should be replaced by $\tau$ in Tab.~\ref{tab:status} and $\tau$ and
$N_\tau$ should be replaced by $z$ and $N_\sigma$ in Eq.~(\ref{eq:fit}),
respectively. We calculate meson propagators using point sources as well as
corner-wall sources, where on a given time slice the source is set to one at the
origin of each $2^3$ cube and zero elsewhere. The use of corner-wall sources reduces
the contribution of higher excited states and thus allows for a more accurate
determination of the screening masses, especially for the positive parity states. 

As stated above, in this study we do not include the contribution from disconnected
diagrams. In the case of charmonium the contribution of disconnected diagrams is
expected to be small, see e.g. Ref.~\cite{Levkova:2010ft}. For $s \bar s$ mesons
disconnected diagrams will cause mixing with the light quark sector in the isospin
zero channel. For vector mesons this mixing is known to be very small and the $\phi$
meson is almost a pure $s \bar s$ state. In SU(3) quark model language this is called
the ideal mixing between SU(3) flavor singlet and SU(3) flavor octet.  Thus,
neglecting the disconnected diagrams seems to be justified also in this case. Mixing
is, however, large for iso-singlet pseudo-scalar mesons. For a realistic study of
$\eta$ and $\eta'$ mesons it is certainly necessary to include contributions from
disconnected diagrams. Thus, we do not pursue a detailed study of the pseudo-scalar
meson correlators in the $s \bar s$ sector.  It is customary, however, to compare the
lattice calculations of the pseudo-scalar meson mass which do not include
disconnected diagrams with the mass of the fictitious un-mixed $s\bar s$ meson that
is estimated using leading order chiral perturbation theory
$m_{\eta_{s\bar{s}}}=\sqrt{2 m_K^2-m_{\pi}^2}=686$ MeV, with $m_K$ and $m_{\pi}$
being the kaon and pion masses, respectively. We will use this approach in what
follows.

Not much is known about the mixing between the light and strange sectors for
iso-scalar mesons in scalar and axial-vector channels. It is expected that there is
strong mixing in the scalar meson sector as well.  The mass of the lowest lying $s
\bar s$ scalar meson considered in our calculation is about $1.12$ GeV as shown in
Fig.~\ref{fig:T=0} (explained below). It is considerably heavier than the lightest
iso-scalar scalar meson $f_0(980)$ but lighter than the next-to-lightest iso-scalar
scalar state $f_0(1370)$.  However, for the axial-vector meson mass we find good
agreement between our calculations and the mass of the $f_1(1420)$ meson, suggesting
that the mixing between the light and strange quark sector is likely to be small in
this case.  Thus, in the strange sector we could study reliably the correlators in
the vector and axial-vector channels.  Moreover, at sufficiently high temperatures we
expect that the contribution of disconnected diagrams will become small because of
screening effects and the weakly interacting nature of the deconfined phase.  Even in
the pseudo-scalar channel our calculations will therefore be reliable for high enough
temperatures.

\section{Zero temperature calculations and determination of charm quark mass}

Before discussing our results on the temperature dependence of the spatial
correlators and the screening masses we need to determine the charm quark mass at the
various values of the gauge coupling used in our finite temperature calculations and
understand the accuracy that is reachable in our approach. For this
purpose we analyze the meson spectrum at zero temperature at five values of the gauge
couplings spread over the range of couplings used in our finite temperature
calculations. The run parameters for these calculations are summarized in
Tab.~\ref{tab:T0}. 

\begin{table}[tb]
\begin{center}
\caption{Gauge coupling ($\beta$), strange ($m_s$) and charm ($m_c$) quark masses,
lattice sizes ($N_\sigma^3\times N_\tau$) and the number of trajectories (traj.) used
for our zero-temperature spectrum calculations.  The light quark mass is fixed
$m_l=m_s/20$.  The meson correlation functions are calculated every 5 (6)
trajectories for $N_\tau=48$ (64).}
\label{tab:T0}
\begin{tabular}{ccccc}
\hline\hline
$\beta$ & $am_s$ & $am_c$ & $N_\sigma^3\times N_\tau$ & traj. \\
\hline
 6.740 & 0.0476 & 0.575 & $48^4$       & 5995 \\
 6.880 & 0.0412 & 0.486 & $48^4$       & 5995 \\
 7.030 & 0.0356 & 0.410 & $48^4$       & 6995 \\
 7.150 & 0.0320 & 0.360 & $48^3\times64$ & 6096 \\
 7.280 & 0.0284 & 0.315 & $48^3\times64$ & 6096 \\
\hline\hline
\end{tabular}
\end{center}
\end{table}

For the determination of the charm quark mass we calculate the masses of $J/\psi$ and
$\eta_c$ mesons for gauge couplings $\beta=10/g^2$ in the interval $[6.39,7.28]$. We
calculate correlation functions at several trial values of the bare charm quark mass
in the range $10\le m_c/m_s\le 14$ using point sources.  We then perform linear
interpolations in the charm quark mass of the spin-averaged charmonium mass,
$(m_{\eta_c} + 3 m_{J/\psi})/4$ and match them to the physical value. This
determines the bare charm quark mass $am_c$ and the quark mass ratio $m_c/m_s$ for
each value of $\beta$.  Finally we fit the $\beta$ dependence of $am_c$ to a
renormalization group inspired ansatz,
\ber
&
\displaystyle
a m_c^{\rm LCP}=\frac{c_0 f(\beta)+c_2 (10/\beta) f^3(\beta)}{1+d_2 (10/\beta) f^2(\beta)}\; ,\\
\label{eq:amc}
&
\displaystyle
f(\beta)=\left( \frac{10 b_0}{\beta} \right)^{-b_1/(2 b_0^2)} \exp\left( -\frac{\beta}{20 b_0}\right)\; ,
\eer
where $b_0$ and $b_1$ are the coefficients of the QCD beta function.  The above
formula defines the line of constant physics for the charm quark mass.  From our fit
we find $c_0=61.0(1.7)$, $c_2=2.76(26)\times 10^5$, and $d_2=3.3(3.7)\times10^2$. The
details of these calculations are presented in Appendix~\ref{ap:mc}.  

We performed calculations of meson correlation functions containing a charm quark in
four different channels corresponding to local meson operators (see
Tab.~\ref{tab:status}) for $\beta=6.74, 6.88, 7.03, 7.15$ and $7.28$ using point and
corner-wall sources.  We extract the zero temperature masses using the ansatz given
in Eq.~(\ref{eq:fit}).  The pseudo-scalar and vector meson masses obtained using the
corner-wall sources are systematically lower by $0.2\%$ compared to the masses
obtained using the point sources. In the case of pseudo-scalar mesons there is also a
small difference of about $0.2\%$ or less between the two tastes corresponding to the
Goldstone and the lightest non-Goldstone modes. For vector mesons we do not see any
statistically significant splitting between states corresponding to different tastes.
However, the largest uncertainty in the value of the vector and pseudo-scalar masses
expressed in physical units arises from the uncertainty in the determination of the
lattice spacing through the calculation of the kaon decay constant. This scale
setting uncertainty is about $1\%$. 

Our determination of scalar and axial-vector meson masses is less accurate due to the
fact that we use the simple two-exponential ansatz given in Eq.~(\ref{eq:fit})
without including higher excited states. As a result the fit results for the meson
masses oscillate as we vary the lower limit $\tau_{min}$ of the fit interval. These
oscillations persist to all values of $\tau_{min}$, where a reasonable signal can be
extracted. To determine the scalar and axial-vector meson masses we average over
results obtained using different fit intervals making sure that $\tau_{min}$ is large
enough that there is no systematic drift in the value of the screening masses beyond
these oscillations.  The typical difference between the averaged value and the
individual fit values of the meson masses is used as an estimate of systematic
errors.

\begin{figure}
\includegraphics[width=8cm]{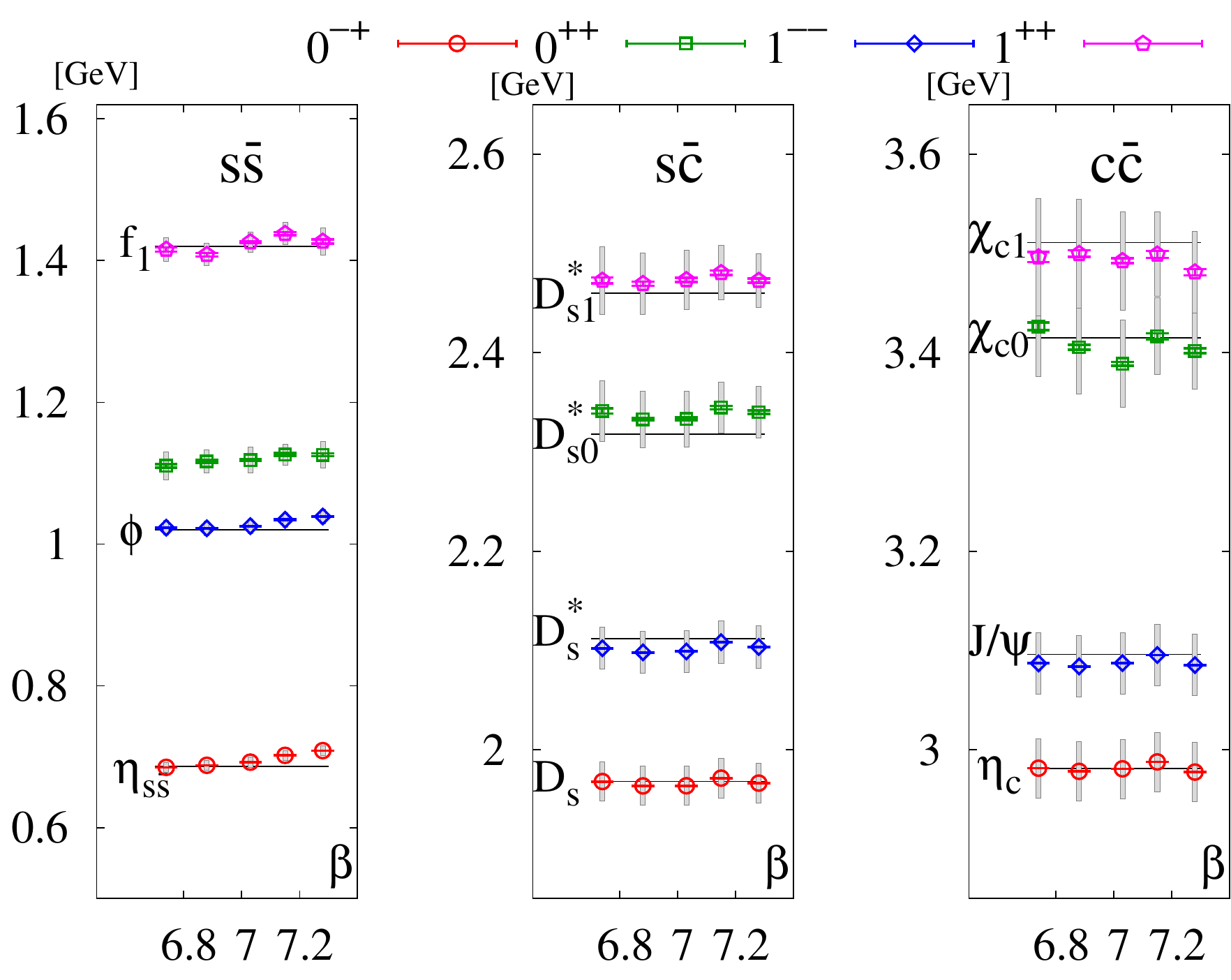}
\caption{Zero-temperature masses of the lowest lying $s\bar s$, $s\bar c$ and $c\bar
c$ mesons.  Corresponding experimental values are depicted by the horizontal lines.
The error bars and broad gray bands indicate the statistical and systematic errors,
respectively.  The systematic errors also include the scale uncertainty (see text).
In $s \bar s$ sector we compare our results with the mass of the un-mixed
$\eta_{s\bar{s}}$ pseudo-scalar meson (see text).}
\label{fig:T=0}
\end{figure}

Our results for the zero temperature masses and comparisons with experimental values
are shown in Fig.~\ref{fig:T=0}. The error bars shown in the figure correspond to the
statistical errors.  The broad gray bands correspond to the systematic errors due to
dependence of our fitted masses on the fit interval and $1\%$ uncertainty of the
scale setting discussed above. As one can see from the figure, our calculations
reproduce the experimental results within the estimated errors.  There is no apparent
cutoff dependence of the meson masses in the beta range studied by us. This is partly
due to the fact that cutoff dependence of the meson masses is compensated by the
cutoff dependence of the kaon decay constant $f_K$ \cite{Bazavov:2012jq} used to set
the lattice spacing. The deviations of the $\phi$ and $\eta_{s\bar{s}}$ meson masses
from the experimental values at the two highest $\beta$ are due to the mistuning of
the strange quark mass. As discussed in Ref. \cite{Bazavov:2014pvz} at $\beta > 7.03$
the values of the strange quark mass are sligltly above the nominal value. After retuning
the valence strange quark mass to the correct value one indeed finds that the experimental
value of the $\phi$ meson mass is reproduced. 

The averaged value of the hyperfine
splitting, $m_{J/\Psi}-m_{\eta_c}$, in our calculations turns out to be $107(1)$ MeV
compared to the experimental value of $113.2(7)$~MeV. The small discrepancy of
$6$~MeV could be due to the missing contributions from disconnected diagrams,
inaccurate tuning of the charm quark mass and slightly larger than the physical light
quark mass. Altogether we find that discretization errors in the charm sector are
considerably smaller than required for the studies at non-zero temperature that we
discuss in the following sections. 

\begin{figure*}[t]
\includegraphics[width=5.8cm]{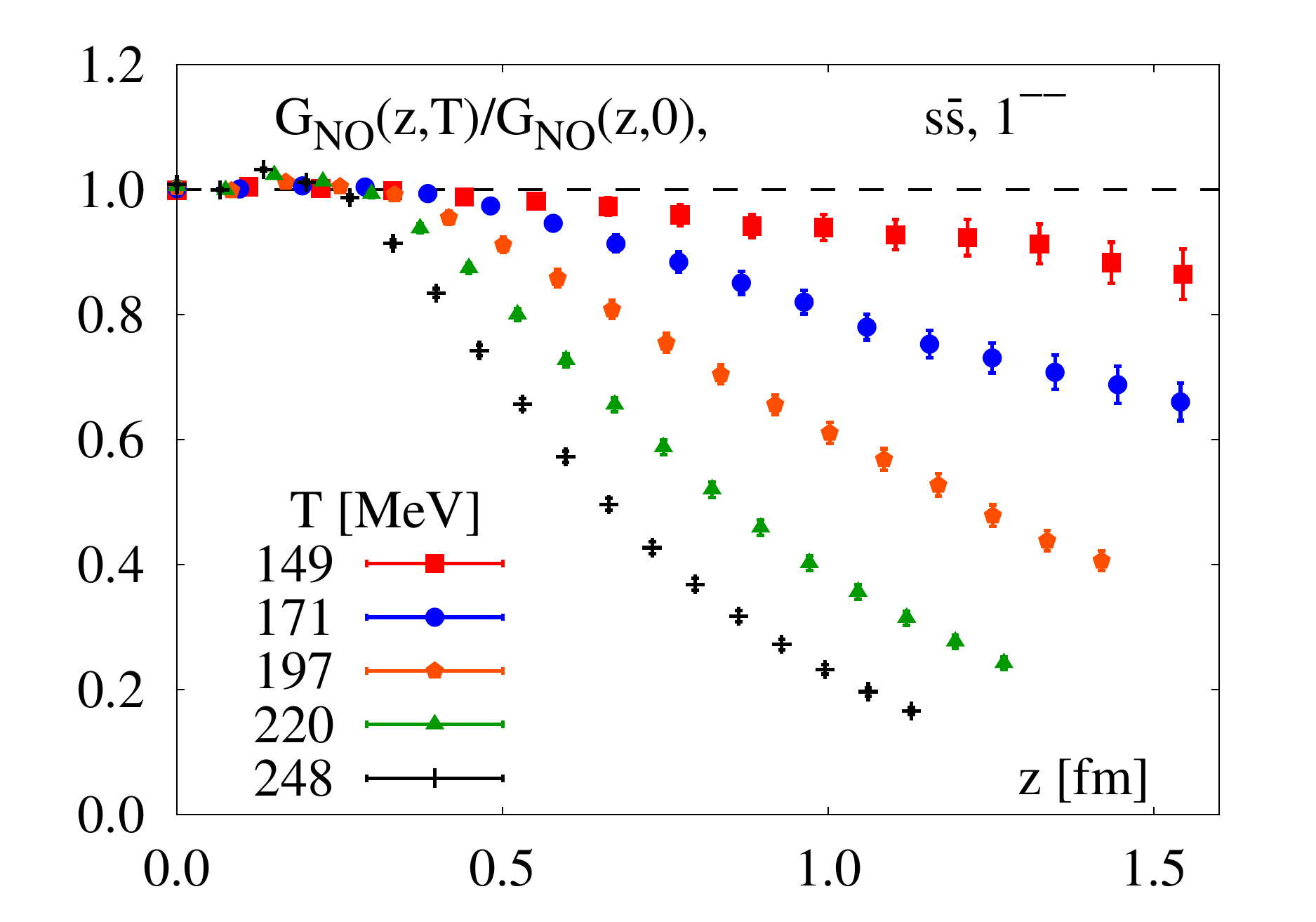}
\includegraphics[width=5.8cm]{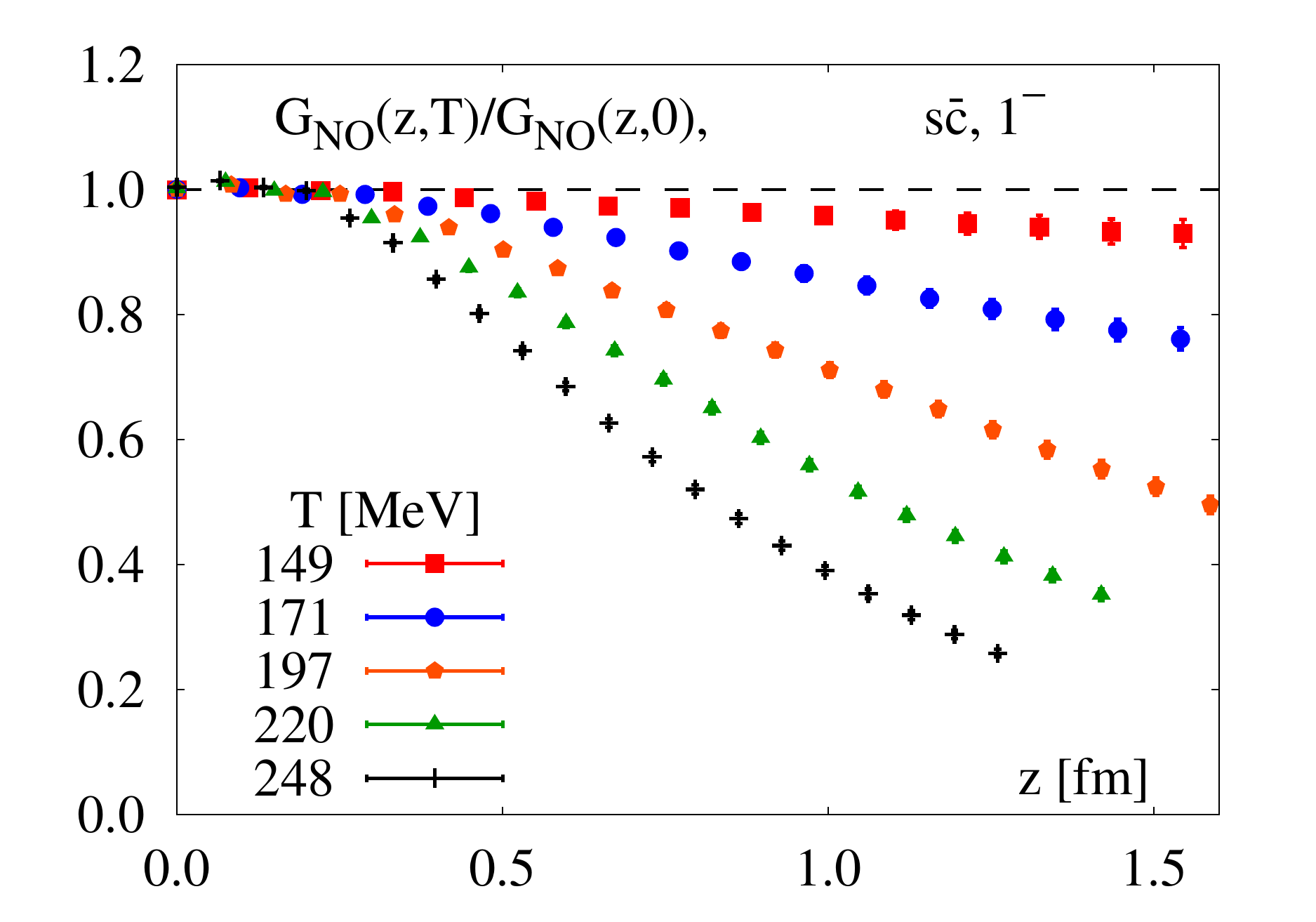}
\includegraphics[width=5.8cm]{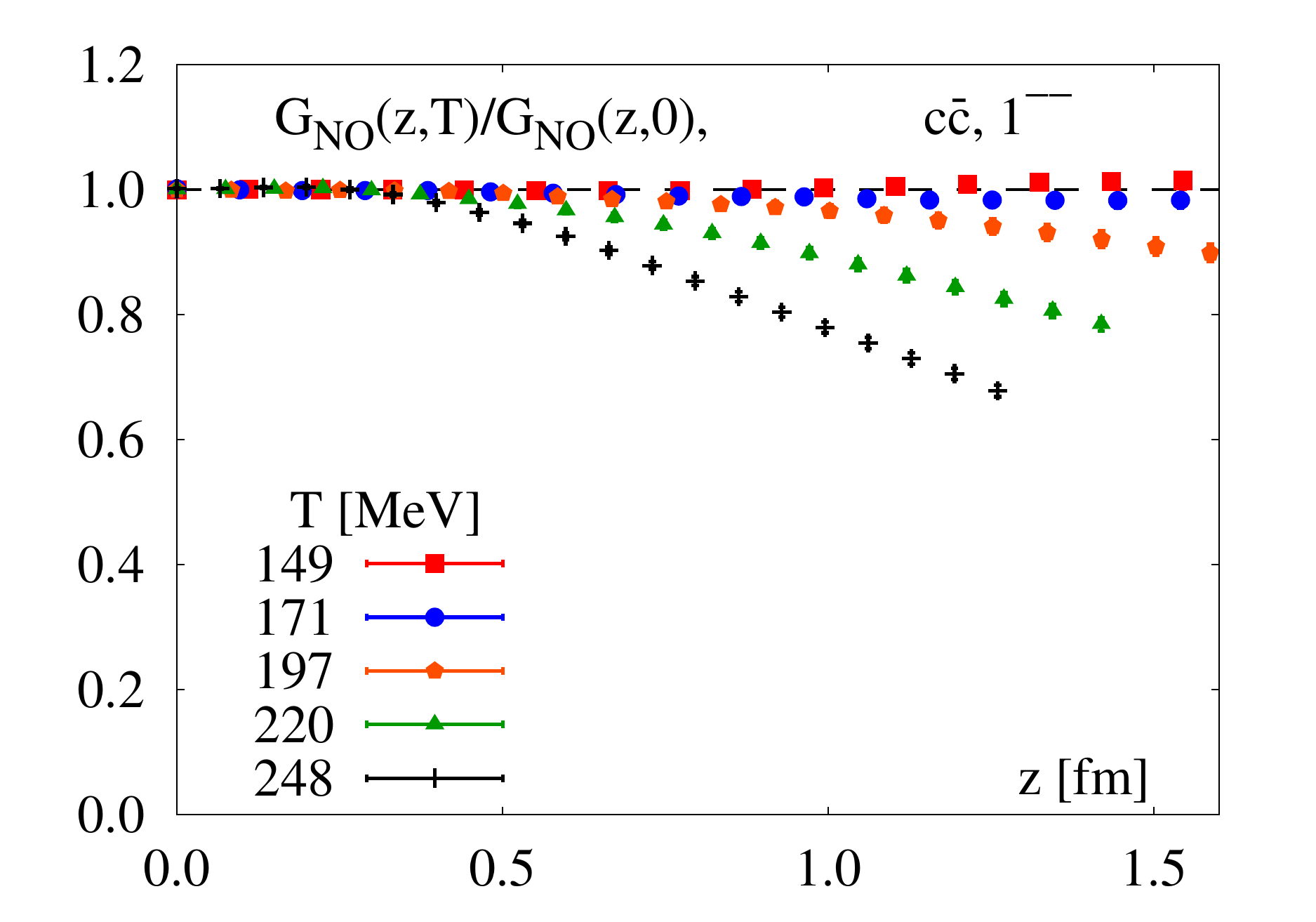}
\caption{Ratios of the non-oscillating (negative parity) part of vector correlators
in $s \bar s$ (left), $s \bar c$ (middle) and $c \bar c$ (right) sectors at different
temperatures to the corresponding zero temperature results. We also show the $J^{PC}$
quantum numbers in each sector. For $D_s$ meson we have $1^-$ as there is no charge
conjugation symmetry (see Tab.~\ref{tab:status}).}
\label{fig:ratn}
\end{figure*}

\section{Temperature dependence of spatial meson correlators}

Having determined the charm quark masses we can perform calculations of the finite
temperature $s\bar{s}$, $s\bar{c}$ and $c\bar{c}$ correlation functions in the four
different quantum number channels that we have analyzed also at zero temperature. 

We start the discussion of our results at non-zero temperature with the temperature
dependence of meson correlators. It was pointed out in Ref.~\cite{Karsch:2012na},
contrary to the temporal correlation functions, spatial correlation functions show
visible changes already in the vicinity of the crossover temperature even in the case
of charmonium.  Since staggered meson correlators in each channel contain both
negative parity (non-oscillating) and positive parity (oscillating) states, it is
important to separate these contributions before studying the temperature dependence
of the correlators. As discussed in detail in Appendix \ref{ap:efms}, it is possible
to define two separate effective correlators for the negative (non-oscillating, $NO$)
and positive (oscillating, $O$) parity states of a staggered meson correlation
function $G(z)$
\begin{subequations}
\label{eq:eff}
\begin{eqnarray}
G_{NO}(z) &\equiv& A_{NO}^2(z) e^{-M_-(z)z} =        \frac{g_1+g_0 x_+}{x_- + x_+} ,\\
G_{O}(z)  &\equiv& A_{O}^2(z) e^{-M_+(z)z}  = (-1)^z \frac{g_1-g_0 x_-}{x_- + x_+} ,
\end{eqnarray}
\end{subequations}
in terms of the local, effective masses, $x_\pm(z)=e^{-M_\pm(z)}$, obtained by
solving the equations
\begin{eqnarray}
Ax_\pm^2\mp Bx_\pm +C=0\; .
\end{eqnarray}
Here, $A=g_1^2-g_2g_0$, $B=g_3g_0-g_2g_1$, $C=g_2^2-g_3g_1$, and $g_i\equiv G(z+i),
i=0,1,2,3$ are the values of the meson correlation function at four successive $z$
values. 
In this description we neglect contribution of periodic boundary to the propagating direction.
Further, one can form ratios of these contributions at different temperatures
to the corresponding zero temperature results. As discussed before, in contrast to
temporal correlation functions, such ratio can directly probe the thermal
modifications of the spectral functions themselves. If there is no change in the
meson spectral functions, these ratios will be equal to one and deviations from unity
will indicate in-medium modification of the meson spectral functions at non-zero
temperature. 

In Fig.~\ref{fig:ratn}, we show the ratio of the negative parity part of the vector
correlator in $s \bar s$, $s \bar c$ and $c \bar c$ channels to the corresponding
zero temperature correlator, obtained with point sources.  At zero temperature, these
non-oscillating parts of vector correlators are dominated by $\phi$, $D_s^{*}$ and
$J/\psi$ states, respectively.  We show the results only for $z/a<18$ in the
$s\bar{s}$ sector and for $z/a<20$ in the $s\bar{c}$ and $c\bar{c}$ sectors. At these
distances the influence of periodic boundary conditions in the spatial directions can
be neglected in the effective meson correlator introduced in Eqs.~(\ref{eq:eff}). This
is discussed in more detail in Appendix \ref{ap:efms}.  In the $s \bar s$ sector, at
large distances, we observe $\sim18\%$ and $\sim38\%$ decrease of this ratio at
$T=149$ MeV and $171$ MeV, respectively. A somewhat smaller but still significant,
$\sim8\%$ and $\sim20\%$ respectively, decrease of this ratio is also seen in the $s
\bar c$ sector. Note that, 
recent lattice studies based on flavor and quantum number correlations 
\cite{Bazavov:2013dta,Bazavov:2014yba} have strongly suggested that open strange and charm
mesons start to melt already around $T_c=(154\pm9)$ MeV. Thus, a $\sim20\%$ deviation
of the in-medium correlator with respect to the vacuum one depicts a thermally
modified spectral function with melted meson state.  For ratios in the $J/\psi$
sector no changes are visible at the lowest $149$ MeV temperature, and even at
$T=171$ MeV the deviations of this ratio from unity are at best a few percent. For
charmonia the deviations of the in-medium correlators, at large distances, with
respect to the vacuum ones become larger than $20\%$ only for $T\gtrsim200$ MeV.  In
all cases, the ratio of correlators decreases with increasing temperature at large
distances. As we will see in the next section this is related to the fact that
screening masses in the negative parity channels increase with respect to their
vacuum values with increasing temperature. 

\begin{figure*}
\includegraphics[width=5.8cm]{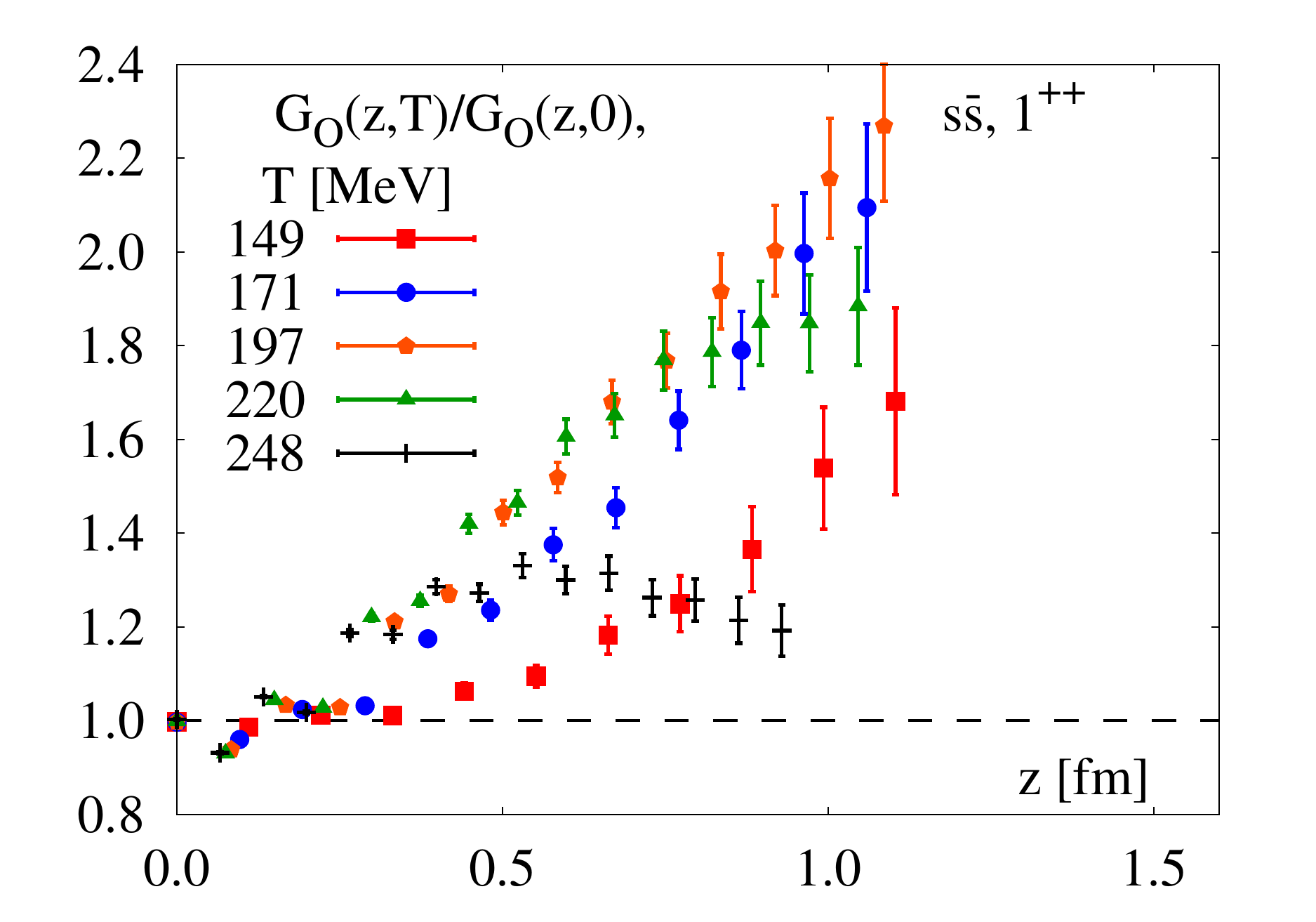}
\includegraphics[width=5.8cm]{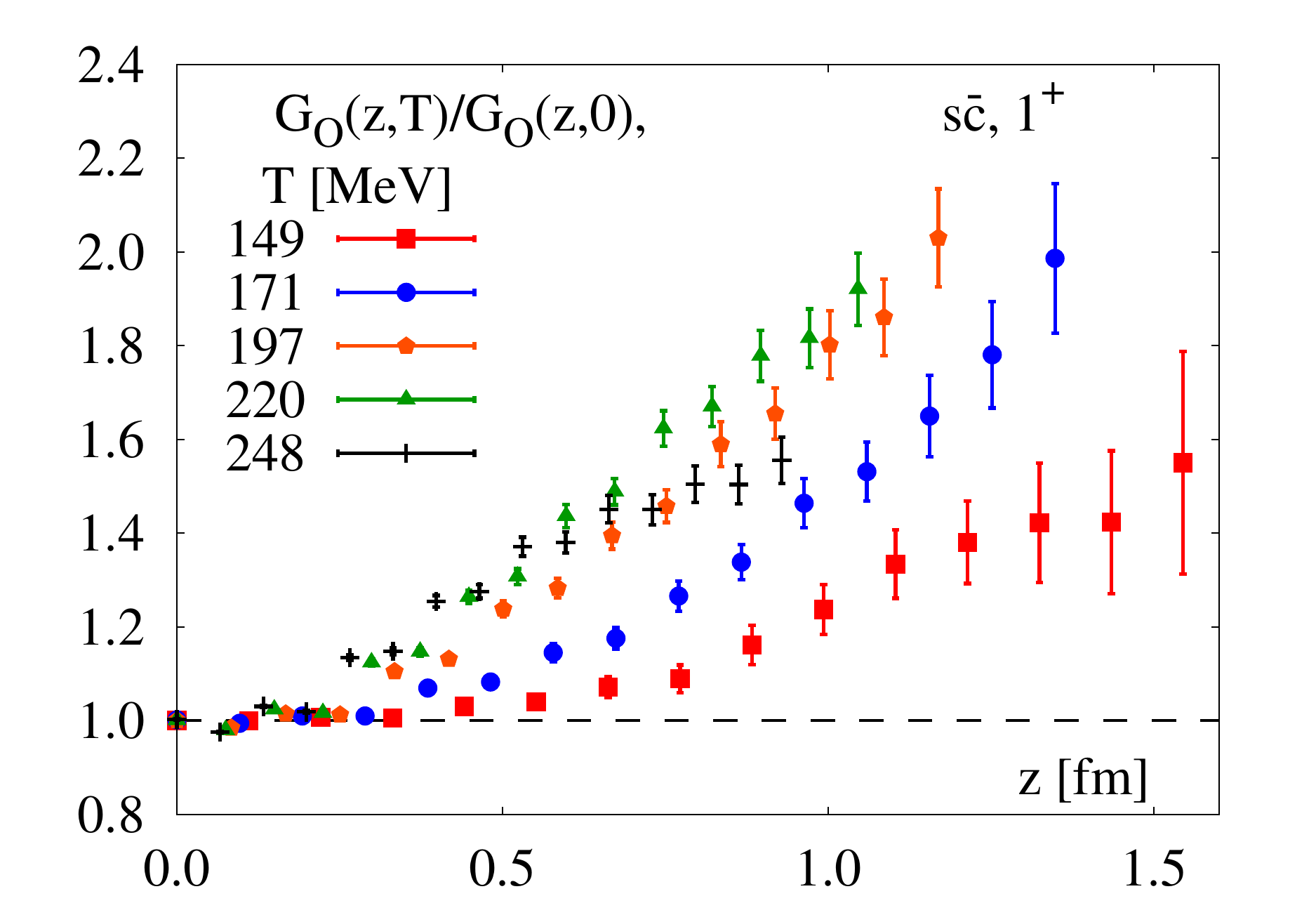}
\includegraphics[width=5.8cm]{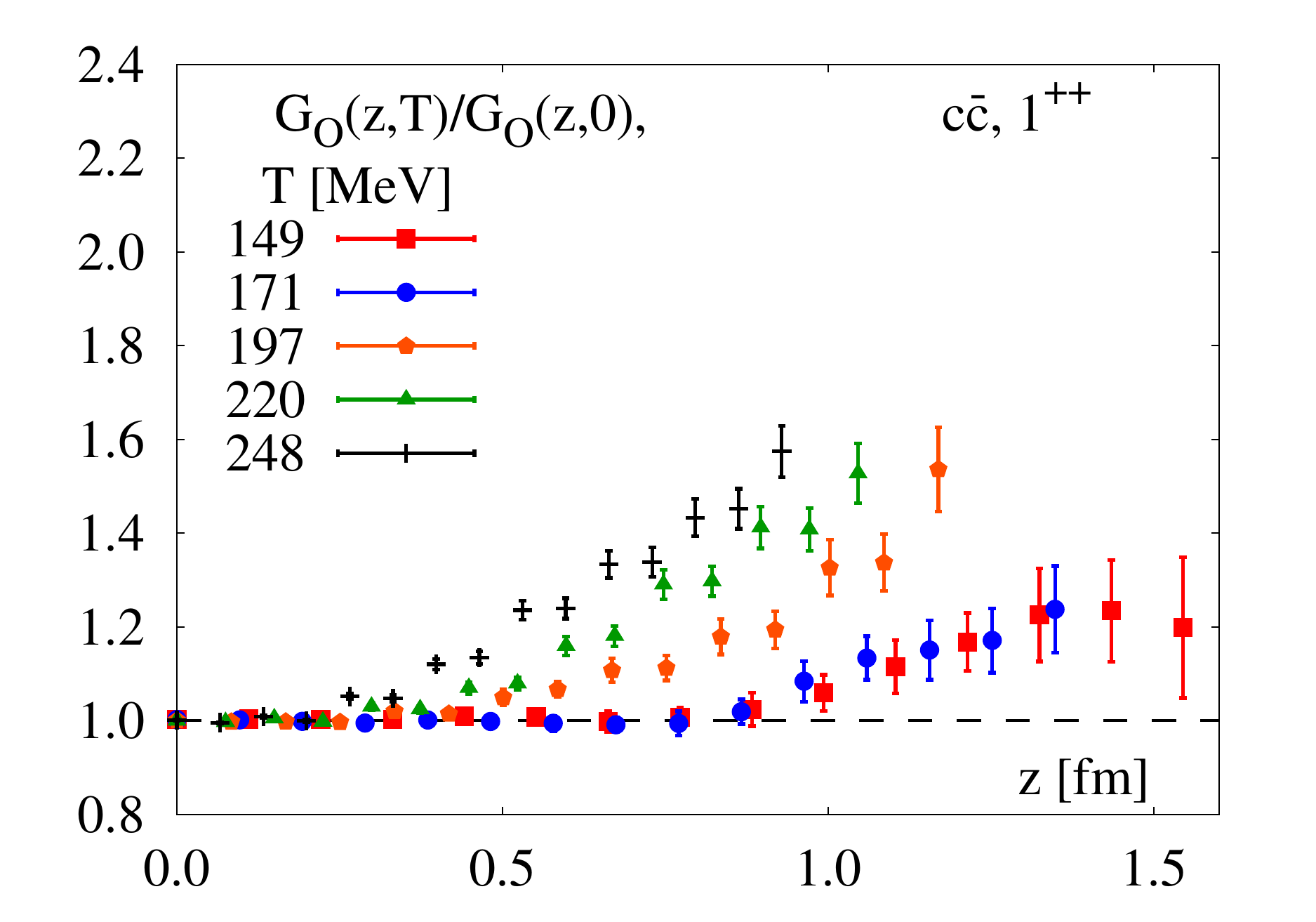}
\caption{Ratios of the oscillating (positive parity) part of axial-vector
correlators in $s \bar s$ (left), $s \bar c$ (middle) and $c \bar c$ (right) sectors
at different temperatures to the corresponding zero temperature results. See Tab.~\ref{tab:status} for $J^{PC}$ assignments.}
\label{fig:ratp}
\end{figure*}

In Fig.~\ref{fig:ratp}, we show the ratio of the positive parity (oscillating) part
of the axial-vector correlator in $s \bar s$, $s \bar c$ and $c \bar c$ channels to
the corresponding zero temperature correlator, again obtained with point sources.  We
only show the ratio for $z/a<15$ as at larger distances the effect of periodic
boundary conditions cannot be neglected when extracting the positive parity
contribution from the correlators (see Appendix B).  The meson states that dominate
the oscillating part of these correlators are $f_1(1420)$, $D_{s1}$ and $\chi_{c1}$.
The ratios of the correlators in this case show a more complex behavior.  At
relatively short distance the ratios of the correlators increases, then depending on
the temperature value and the quark content the ratio can also decrease both as
function of $z$ and the temperature.  As will be discussed in the next section, this
feature of the correlator ratios is closely related to the behavior of screening
masses of the positive parity states. For not too large temperatures the positive
parity screening masses decrease compared to their vacuum values.  This corresponds
to the increase in the ratio of the correlators at large $z$. At sufficiently high
temperature the screening masses start to increase again, which then leads to the
decrease in the ratio of the positive parity correlators. This tendency is seen in
the $s\bar s$ and $s \bar c$ sectors in Fig. ~\ref{fig:ratp}.

Similar to the case of the negative parity states the size of the medium
modification varies with the heavy quark content.  It is the largest for $s \bar s$
mesons and is the smallest for $c \bar c$ mesons. In fact, for charmonium the ratio of
the correlators is equal to one for $z < 1$ fm and $T< 171$ MeV, while for the other
two cases it is significantly above one already at the lowest temperature.
Furthermore, the size of medium modifications of the correlator ratio in $c \bar c$
sector is much larger than for the negative parity part of the vector correlator.
This is what one would expect in sequential charmonium melting picture, where the
larger and more loosely bound $\chi_c$ states dissolve at lower temperature than
$J/\psi$. 

\begin{figure*}
\includegraphics[width=6.8cm]{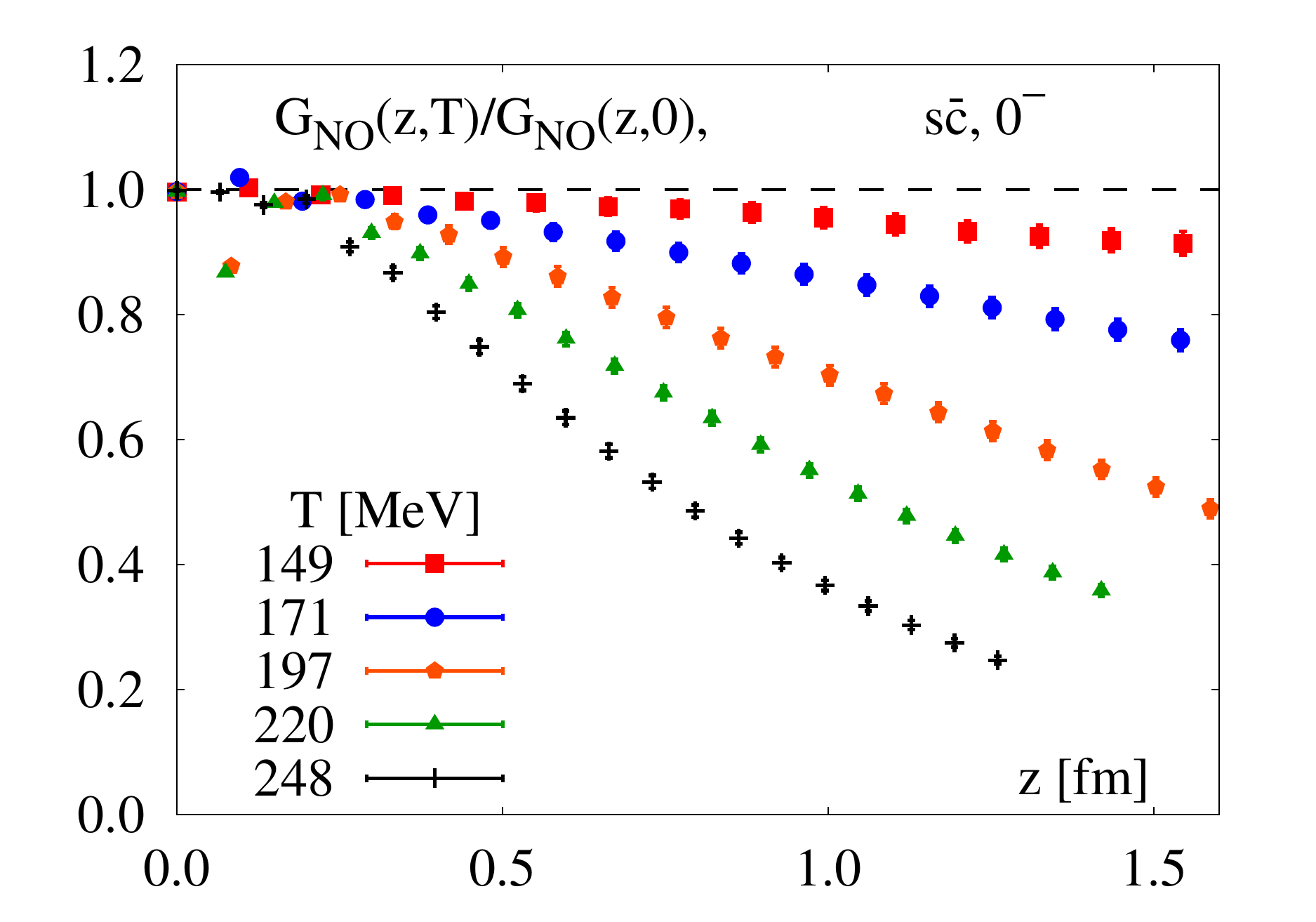}
\includegraphics[width=6.8cm]{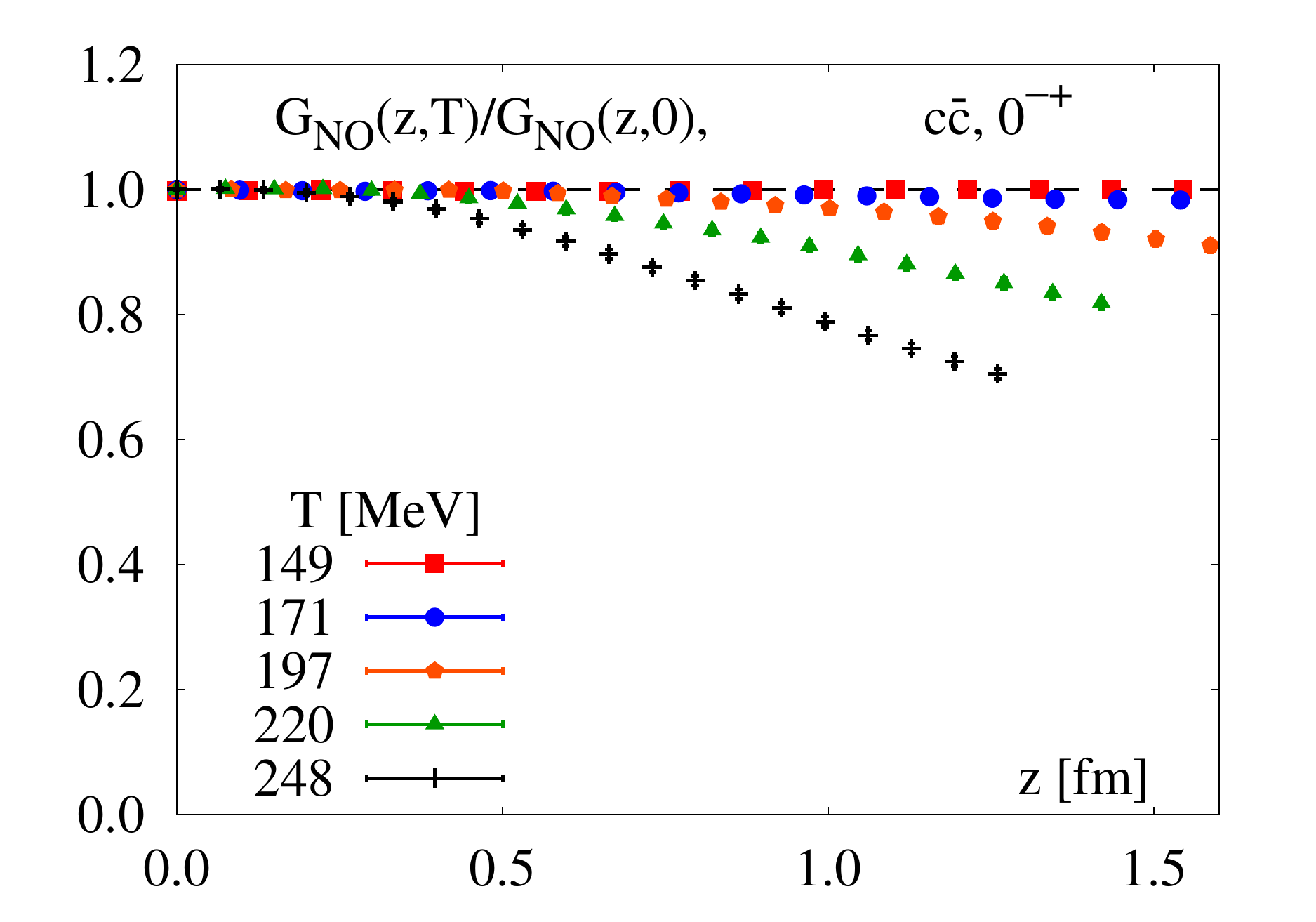}
\includegraphics[width=6.8cm]{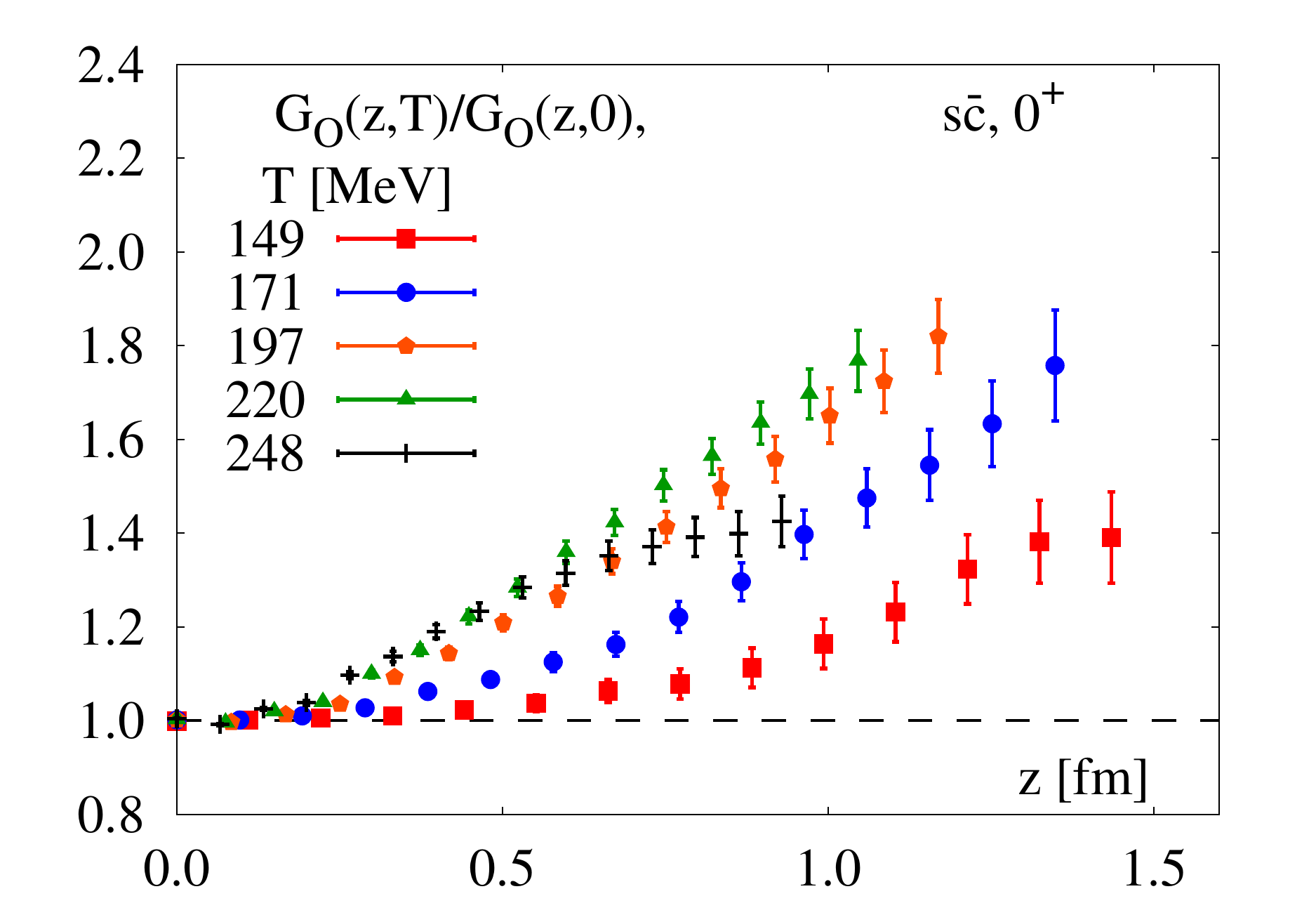}
\includegraphics[width=6.8cm]{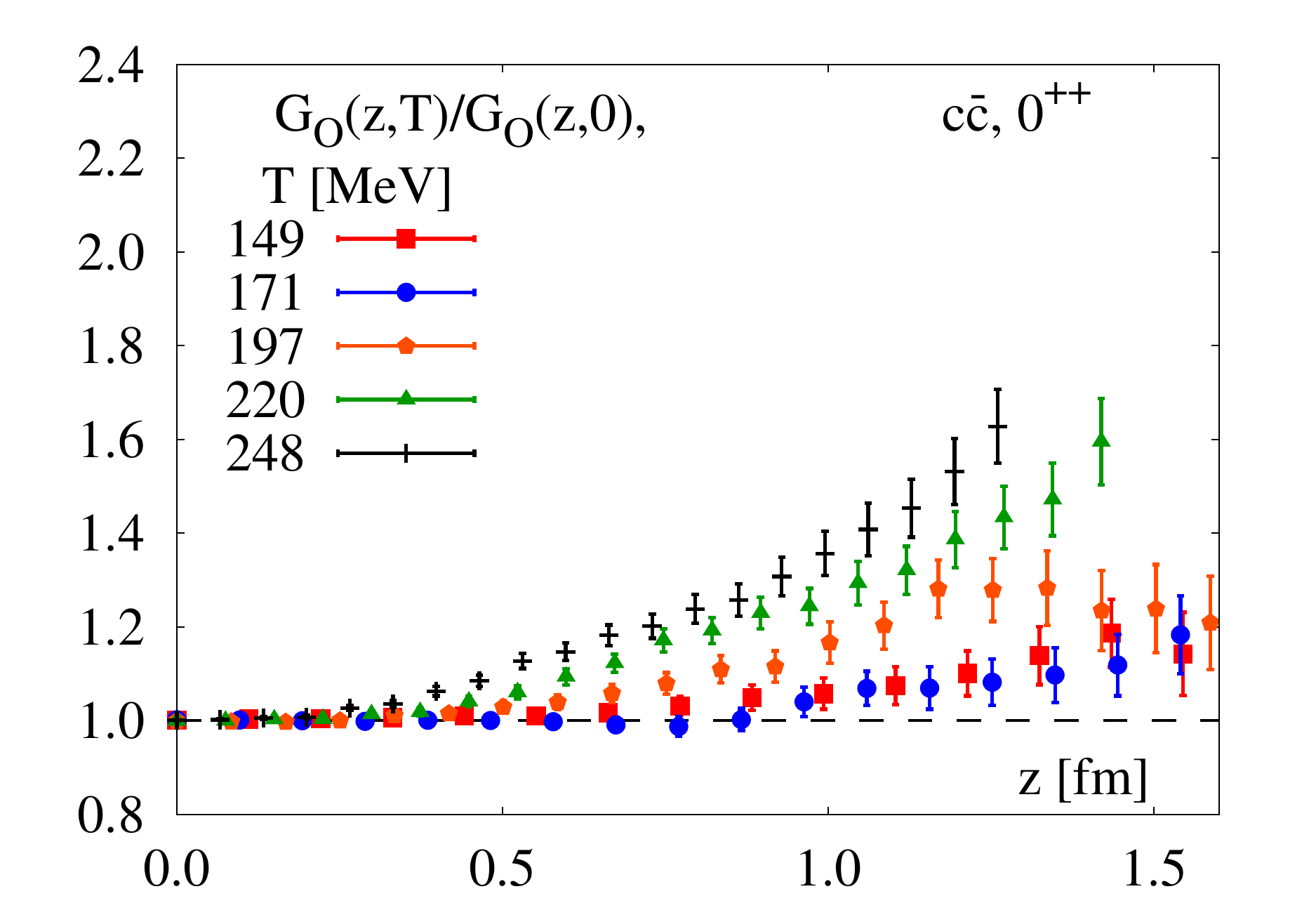}
\caption{Ratios of the non-oscillating (negative parity) part of the pseudo-scalar
correlators (top) and oscillating (positive parity) part of scalar correlators
(bottom) in $s \bar c$ (left), and $c \bar c$ (right) sectors at different
temperatures to the corresponding zero temperature results.}
\label{fig:rat_sps}
\end{figure*}

For $s \bar c$ and $c \bar c$ sectors we also consider the pseudo-scalar and scalar
correlators, as the contributions of the disconnected diagrams are absent for the
former and small for the later. The results are shown in Fig.~\ref{fig:rat_sps}. The
ratios of the non-oscillating part of the pseudo-scalar correlators to the
corresponding zero temperature results are shown in the upper panels of Fig.~\ref{fig:rat_sps}. For small temperatures the lowest states that dominate the
pseudo-scalar correlators are $D_s$ and $\eta_c$ mesons.  The temperature dependence
of these ratios is similar to the ratios of the non-oscillating part in the vector
channel, i.e. they decrease monotonically with increasing temperatures. For $c\bar c$
case the ratio only shows significant modifications for $T\gtrsim200$ MeV, while for
$s \bar c$ sizable modifications are visible already at the lowest temperature. The
ratio of the positive parity contribution is shown in the lower panels of 
Fig.~\ref{fig:rat_sps}, where corresponding states at $T=0$ are $D^\ast_{s0}$ and $\chi_{c0}$ for $s\bar{c}$ and $c\bar{c}$ channels, respectively.
The ratio becomes larger than one. In $c\bar c$ sector the deviations of this
ratio from one are larger than those in the pseudo-scalar case and set in at lower
temperatures. Again, these results are suggestive of sequential melting of the
charmonia states. The size of medium modifications of the correlator ratio also
depends on the heavy quark content: it is larger in the $s \bar c$ sector than in $c
\bar c$ sector.

The ratio of spatial charmonium correlators and corresponding zero temperature
correlators was first studied in Ref.~\cite{Karsch:2012na} in the pseudo-scalar
channel. Strong in-medium modifications of this ratio were found for $z T>1$.  The
magnitude of medium effects in the ratios of pseudo-scalar correlators calculated
with the HISQ action is similar to those obtained previously with the p4 action at
the same value of $T/T_c$.  The large changes in the ratio of the spatial
correlators, that could be indicative of significant in-medium modification or
dissolution of $1S$ charmonium states, are in contrast to the very mild temperature
dependence of the analogous ratios of the reconstructed temporal correlators. 
The reason for this apparent difference lies in the fact that at non-zero temperature 
the temporal correlators are defined only at relatively small separations
$\tau <1/(2T)$ and, thus have a limited sensitivity to the in-medium modifications
of the spectral functions. As one can see from Figs.~\ref{fig:ratn}--\ref{fig:rat_sps} the
medium modification of the spatial correlators for $z <1/(2T)$ is also quite
small. As discussed in the section I the access to larger separation is the main
reason why the spatial correlation functions are more sensitive to the in-medium
modification of the spectral functions.

\section{Large distance behavior of spatial meson correlators and screening masses}

We fit the large distance behavior of the spatial correlators using
Eq.~(\ref{eq:fit}) and extract screening masses in various channels for $s\bar s$, $s
\bar c$ and $c \bar c$ mesons.  In our study of PS and V screening masses we use
point and corner-wall sources. For $c \bar c$ sector the differences between the
results obtained using point and corner-wall sources are quite small. It is typically
around 0.4\%, except for the three highest temperature, where it reaches 3\%.
Similarly in the $s \bar c$ sector, the difference between point and corner-wall
source results is typically about 1\% for all temperatures except the three highest
ones, where it is 3\%. Larger differences between point and corner-wall source
results are seen in the $s \bar s$ sector, where they reach 3\% at the highest five
temperatures, and are about 1\% at other temperatures.  In the following we will use
the results from corner-wall sources when presenting our results on the screening
masses in the PS and V channels. In the S and AV channels screening masses can be
reliably extracted only by using the corner-wall sources.  The effects of taste
symmetry breaking are only visible for the negative parity states in PS and S
channels at low temperatures, where they are about 1.5\%. For temperature above 200
MeV we do not find any statistically significant effect of taste splitting.

\begin{figure}
\includegraphics[width=8.0cm]{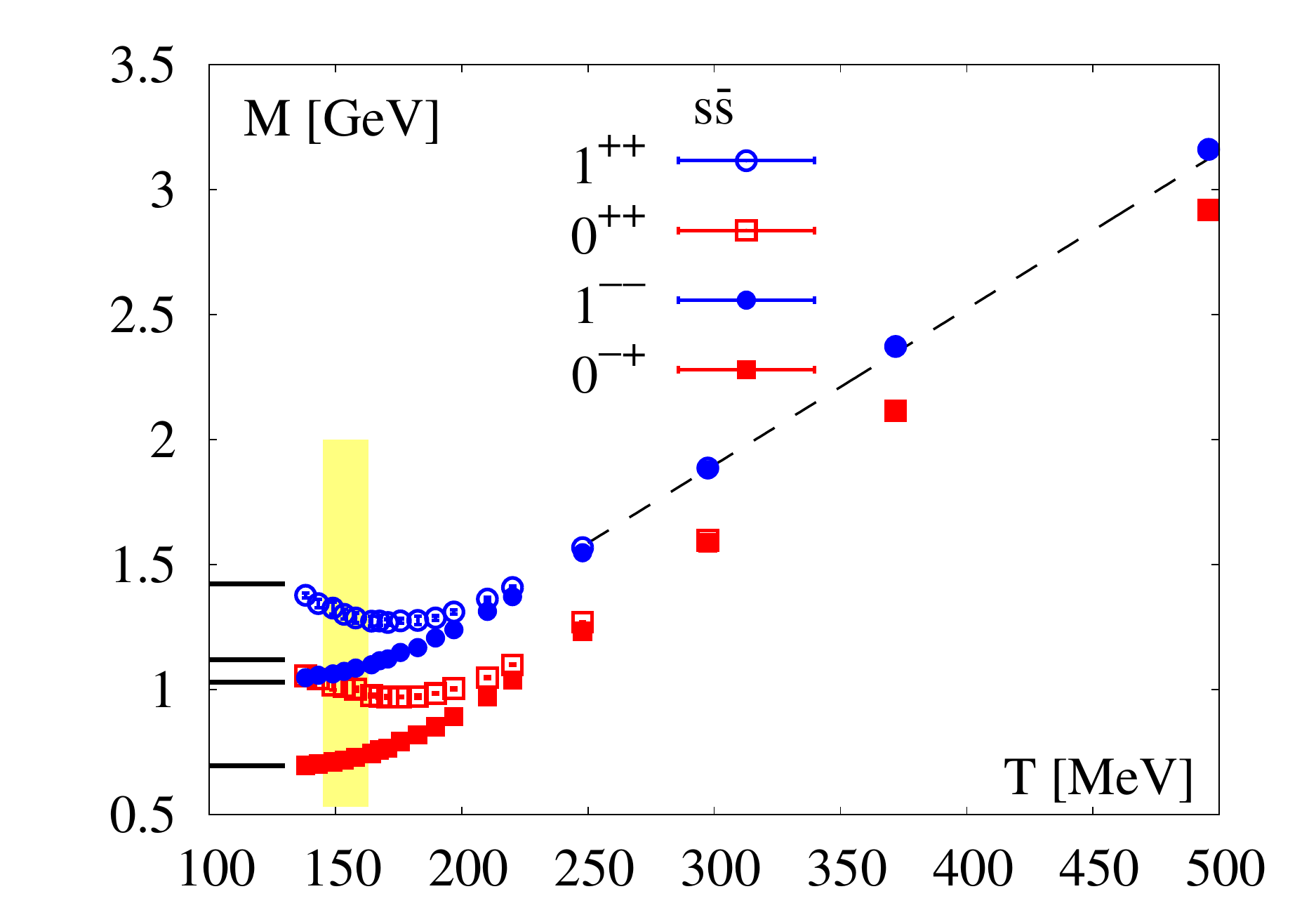}
\includegraphics[width=8.0cm]{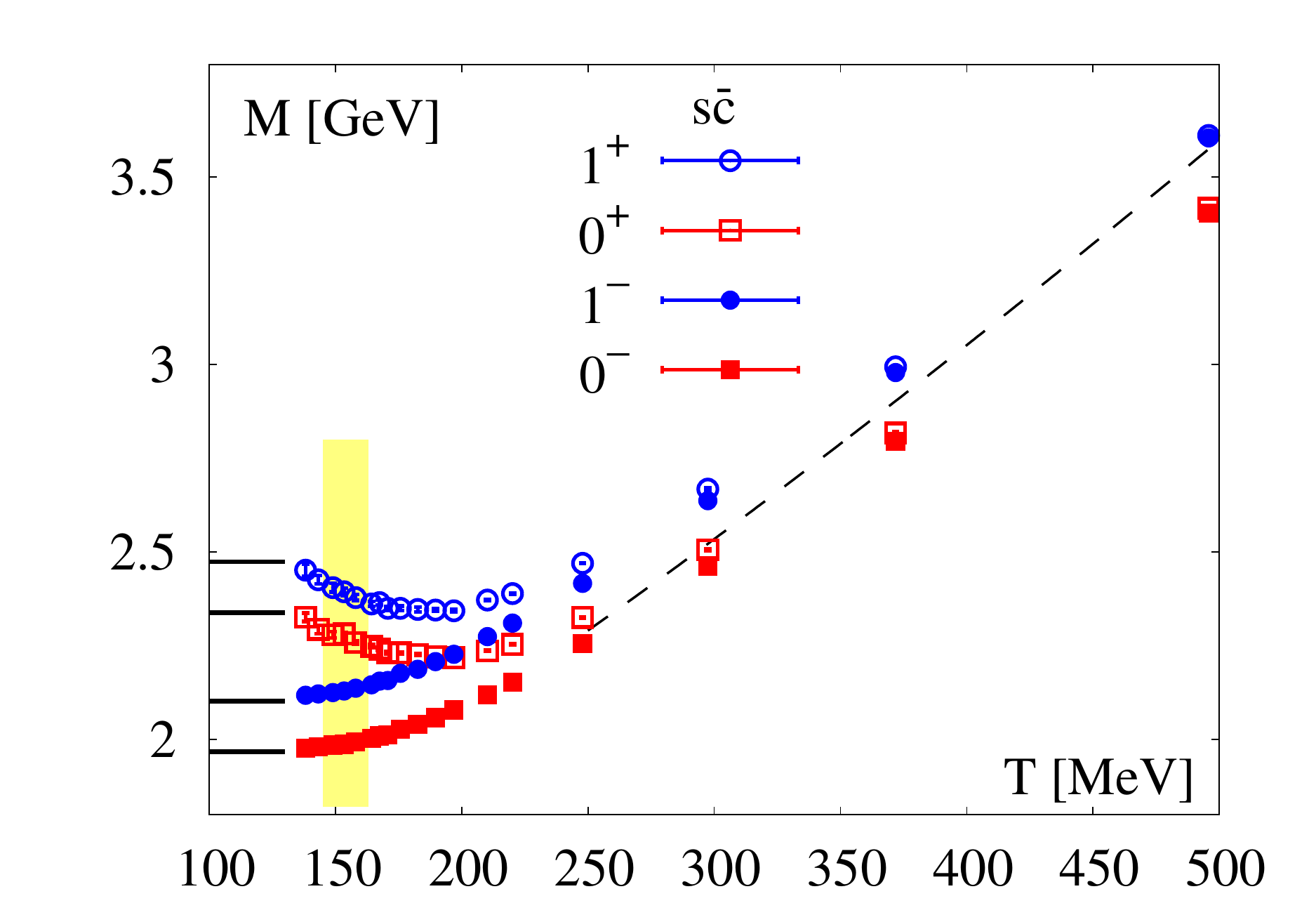}
\includegraphics[width=8.0cm]{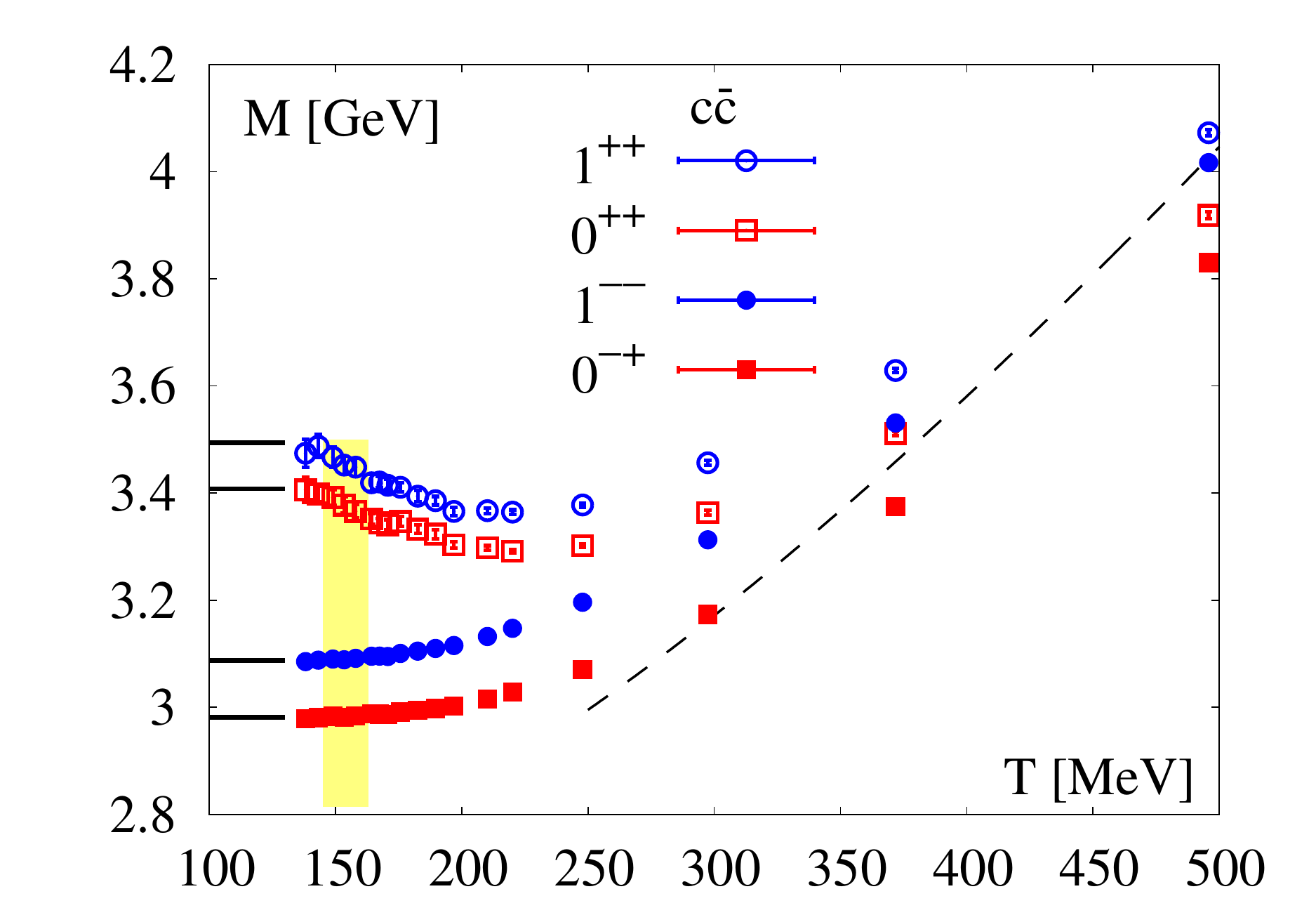}
\caption{Screening masses for different channels in $s\bar s$ (top), $s\bar c$
(middle) and $c \bar c$ (bottom) sectors as functions of the temperature. The solid
horizontal lines on the left depict the corresponding zero temperature meson masses.
The shaded regions indicate the chiral crossover temperature $T_c=(154\pm9)$ MeV.
The dashed lines are the corresponding free field theory result (see text).}
\label{fig:mscr}
\end{figure}

All our results on screening masses in the $s \bar s$, $s\bar c$ and $c\bar c$
sectors are summarized in Tabs.~\ref{tab:app2ss}--\ref{tab:app2cc}, respectively,
given in Appendix~\ref{ap:cal} and are shown in Fig.~\ref{fig:mscr}.  The error bars
in the figure indicate the statistical and systematic errors added in quadrature.  We
expect that at very high temperature the screening masses are given by
Eq.~(\ref{eq:Mfree}).  We show the free theory (leading order perturbative) results
as dashed lines in Fig.~\ref{fig:mscr}. For this we need to specify the quark masses.
The quark masses depend on the renormalization scale which is not specified at
leading order. A natural choice of the renormalization scale would be to identify it
with the lowest scale in the problem provided that this lowest scale is still in the
perturbative region. In our case there are two relevant energy scales, the charm
quark mass $m_c$, and the thermal scale which here is taken to be $2 \pi T$. For
temperatures $T>200$~MeV both scales are comparable. Therefore, we could take the
charm quark mass as the renormalization scale and the corresponding value
$m_c(\bar{\mu}=m_c)=1.275$~GeV from the Particle Data Group \cite{Beringer:1900zz}.  Using the
renormalization invariant ratio of charm to strange quark mass $m_c/m_s=11.85$
\cite{Davies:2009ih} and the above value, we can determine the value of the strange
quark mass at the same renormalization scale to be $m_s=0.108$ GeV.  This completely
specifies our free theory prediction.

As one can see from Fig.~\ref{fig:mscr}, there are three distinct regions: the low
temperature region, where the screening masses are close to the corresponding vacuum
masses (solid lines), the intermediate temperature region, where we see significant
changes in the value of the screening masses with respect to the corresponding vacuum
masses, and finally, the high temperature region, where the screening masses are
close to the free theory result (dashed lines).  In the high temperature region,
there clearly are no meson bound states anymore.  The onset of the high temperature
behavior is different in different sectors.  In the $s \bar s$ sector it starts at
around $T=210$ MeV.  In the $s \bar c$ sector it starts at $T=250$ MeV, while in $c
\bar c$ sector it starts at $T>300$ MeV.  As the temperature increases, we see that
the screening masses corresponding to negative parity states increase monotonically,
while the screening masses in the positive parity states first decrease before
starting to rise towards the asymptotic high temperature values. In the intermediate
temperature region the screening masses of opposite parity partners start to approach
each other and we observe a significant rearrangement of the ordering of screening
masses in different channels.  At sufficiently high temperatures the PS and S
($0^{-+}$ and $0^{++}$) as well as V and AV ($1^{--}$ and $1^{++}$) screening masses
become degenerate.  In the $s \bar s$ sector this is evident for $T>220$ MeV, while
for the two other sectors it happens at higher temperatures due to the larger
explicit breaking of parity by the charm quark mass. In the high temperature region
the screening masses in the PS channel are smaller than the screening masses in the V
channel. This behavior has been observed previously in lattice calculations
\cite{Cheng:2010fe} and in calculations using Dyson-Schwinger equations
\cite{Wang:2013wk}.

\begin{figure}
\includegraphics[width=8.0cm]{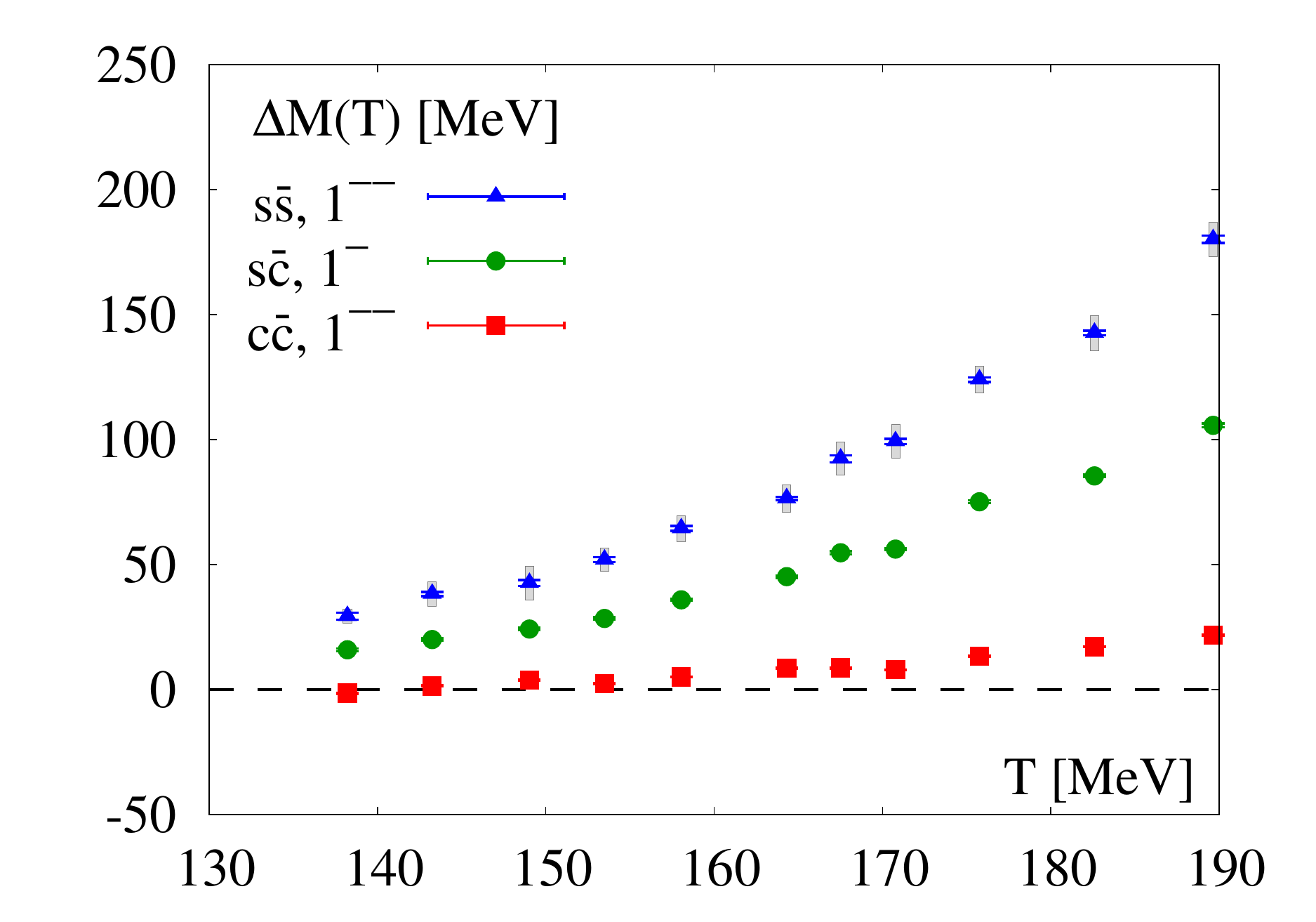}
\includegraphics[width=8.0cm]{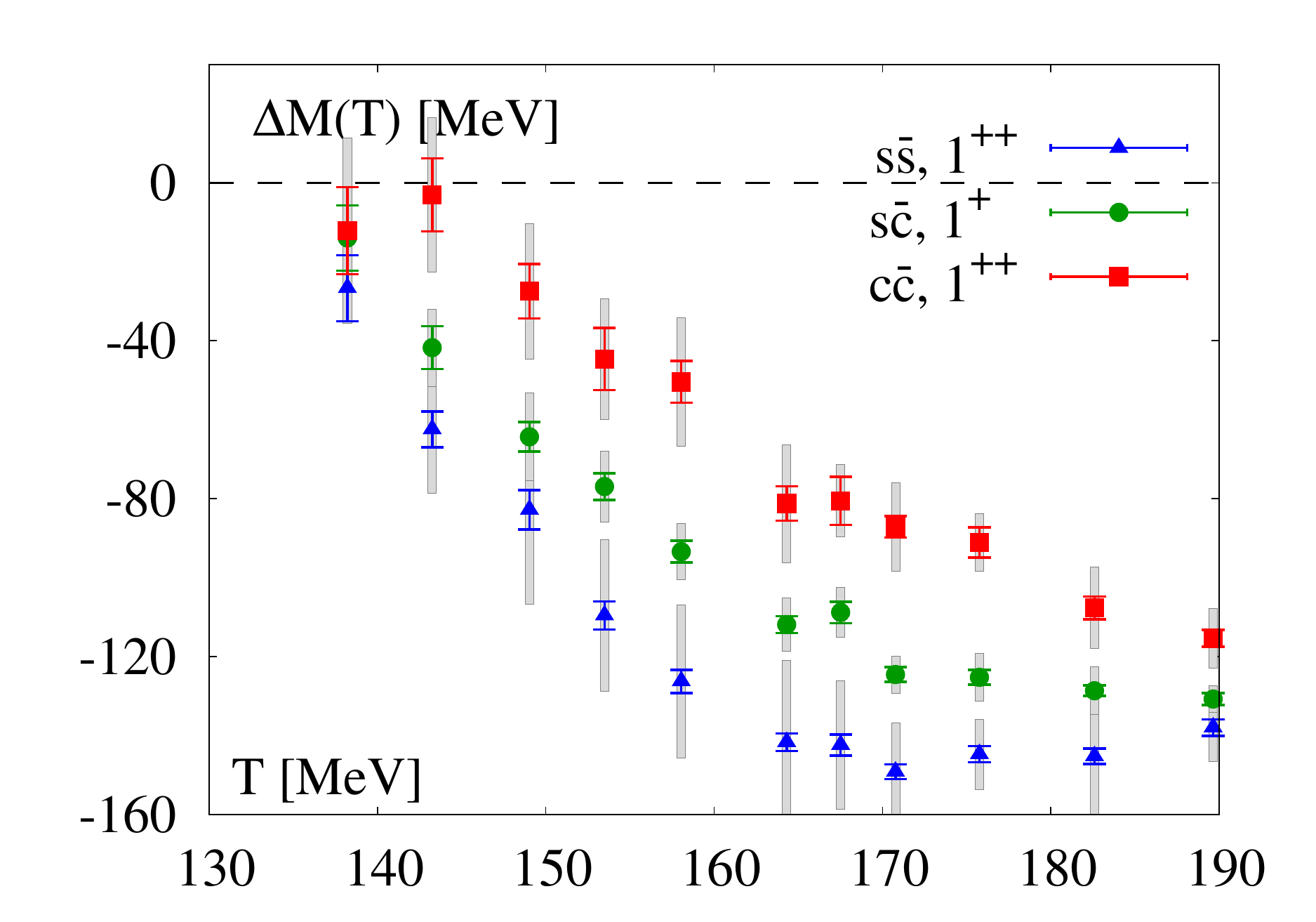}
\caption{The difference, $\Delta M(T)=M(T)-m_0$, of the screening masses, $M(T)$, and
the corresponding vacuum masses, $m_0$, for the negative (top) and positive (bottom)
parity states as functions of the temperature.}
\label{fig:deltaE}
\end{figure}

In order to emphasize the different behavior of negative and positive parity
screening masses in the low and intermediate temperature regions it is convenient to
consider the difference between the screening mass and the corresponding vacuum
masses $m_0$ calculated at $T=0$ 
\begin{equation}
\Delta M(T)=M(T)-m_0\; . 
\end{equation}
It is tempting to interpret this difference as the change in the binding energy of
meson states, however, the relation between the screening mass and the pole mass only
holds as long as there is a well defined bound state. Nonetheless, $\Delta M$ could
provide some constraints on the change of the binding energy in the low and
intermediate temperature  regions.  We show our results for $\Delta M(T)$ for vector
($1^-$) and axial-vector ($1^+$) $s\bar s$, $s\bar c$ and $c \bar c$ mesons in
Fig.~\ref{fig:deltaE}.  The error bars and gray bands indicate the statistical and
systematic errors, respectively.  In all cases $\Delta M$ increases for negative
parity states and decreases for positive parity states in the considered temperature
region.  This corresponds, of course, directly to the pattern seen in the behavior of
ratios of spatial correlation functions (see Figs. \ref{fig:ratn}, \ref{fig:ratp} and
\ref{fig:rat_sps}).  At higher temperature the positive parity screening masses will
start increasing again (see Fig.~\ref{fig:mscr}) leading to the non-monotonic
behavior of the correlator ratios in the $s \bar s$ and $s \bar c$ sectors.  Except
for the negative parity (S-wave)  charmonium states we clearly see that in all other
cases in-medium modifications lead to significant deviations of screening masses from
pole masses already in the crossover region. In the case of $s\bar{s}$ states this is
the case even below $T_c$.  For the S-wave charmonium states screening and pole
masses are nearly compatible up to temperatures of about $200$ MeV. This is
consistent with the small deviations from unity observed for ratios of zero and
finite temperature correlators in these quantum number channels and may indicate that
these states do persist as bound states at least up to this value of the temperature.
Thus, in the charmonium case the temperature dependence of $\Delta M$ provides some
hints for sequential thermal modification: It shows a strong decrease of screening
masses starting in the crossover region for scalar and axial-vector channels
corresponding to 1P charmonium states ($\chi_{c0}$ and $\chi_{c1}$) and very little
change in the pseudo-scalar and vector channels corresponding to 1S charmonium states
($\eta_c$ and $J/\psi$).

\begin{figure}
\includegraphics[width=8.5cm]{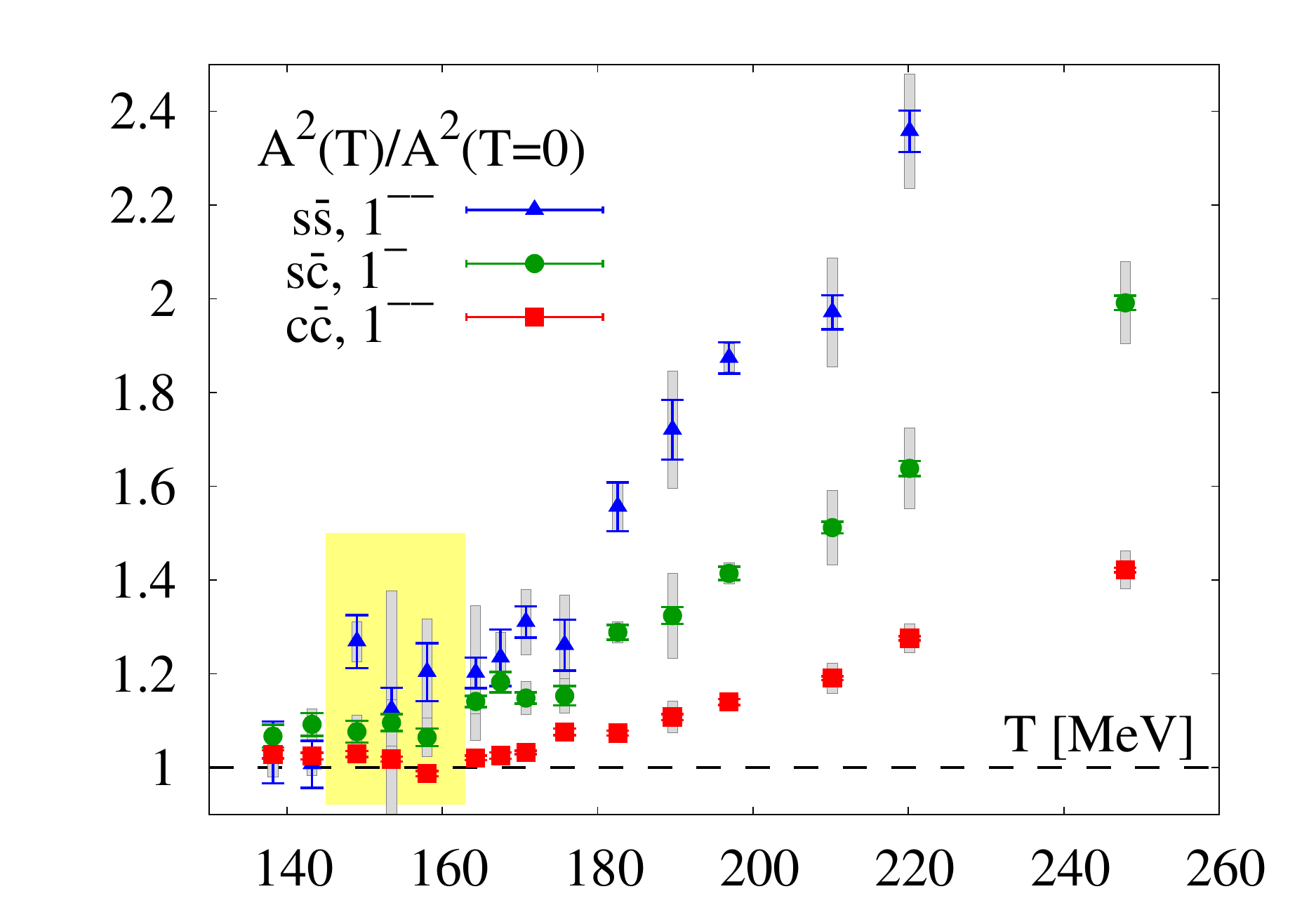}
\caption{The ratio of the amplitude $A_{NO}^2$ (see Eq.~(\ref{eq:fit})) of the
spatial correlators to the corresponding zero temperature amplitude in the vector
channel as function of the temperature.  The shaded region indicates the chiral
crossover transition.}
\label{fig:amp}
\end{figure}

\begin{figure}
\includegraphics[width=8.5cm]{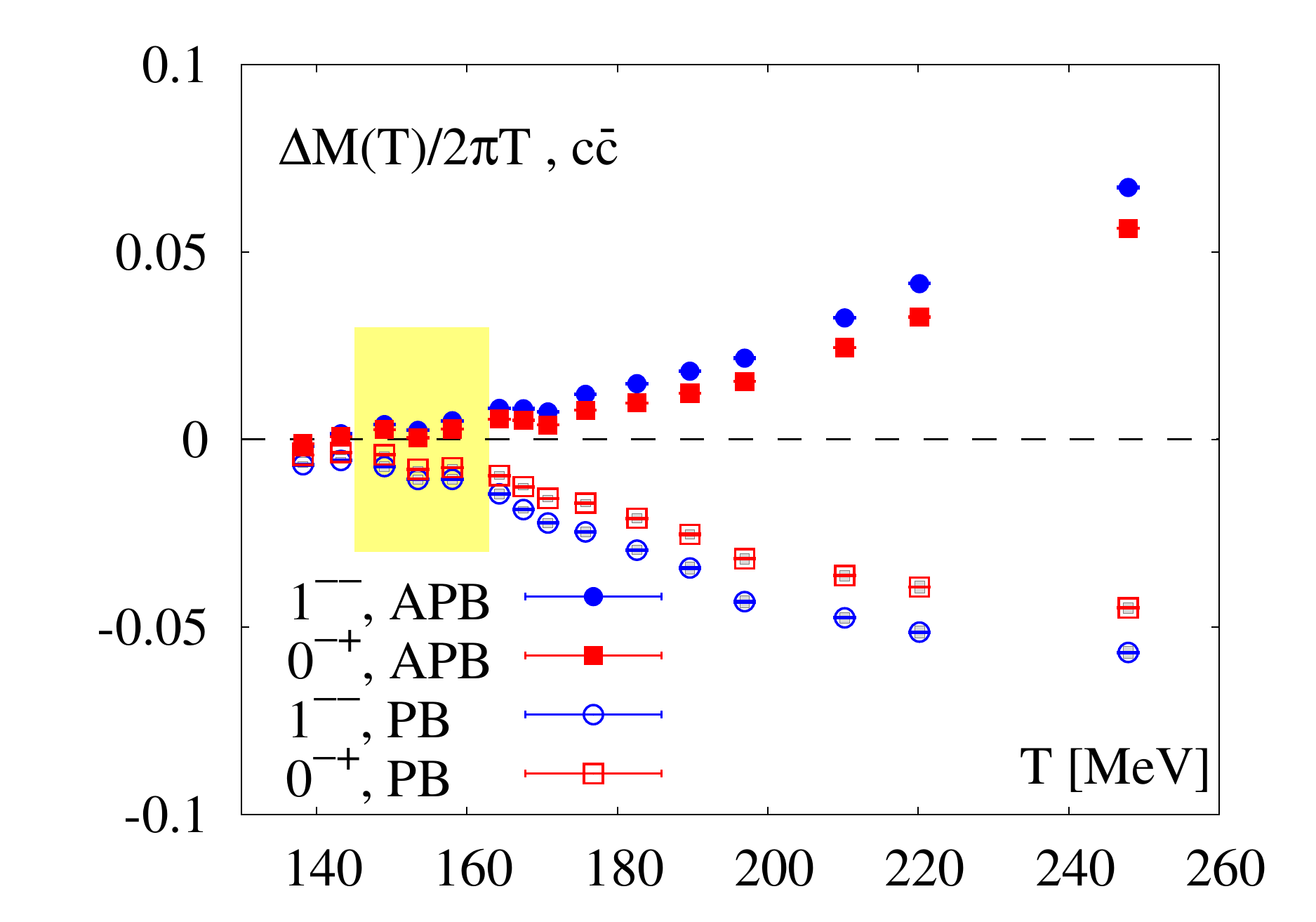}
\caption{The difference, $\Delta M = M(T)-m_0$ normalized by $2\pi T$, between the
screening masses and the corresponding zero temperature masses for the pseudo-scalar
($0^{-+}$) and vector ($1^{--}$) channels, by imposing usual anti-periodic (filled
symbols) as well as periodic (open symbols) temporal boundary conditions for the
fermions (see text). The shaded region shows the chiral crossover transition.}
\label{fig:pb}
\end{figure}

The amplitudes $A_{NO}^2$ and $A_O^2$ appearing in Eq.~(\ref{eq:fit}) are related to
the wave function of meson states and to the meson decay constants (corresponding to weak decays
or a decay into a virtual phtoton) in the zero temperature limit. 
In particular, for point sources and mesons consisting of heavy quarks they are proportional to the square of
the wave function at the origin and the derivative of the square of the wave function
at the origin. As we are interested in signatures for the melting of meson states at
high temperature, it is worth studying the temperature dependence of these
amplitudes. In Fig.~\ref{fig:amp} we show the ratios of the amplitude $A_{NO}^2$ for
the spatial meson correlators in the vector channel to the corresponding zero
temperature result. If meson states exist in the medium with little or no
modifications this ratio should be close to one. For the strange and strange-charm
mesons we see that there are deviations of this ratio from one already at relatively
low temperatures and these deviations are increasing with increasing temperature. For
charmonium the ratio of the amplitude to the zero temperature amplitude is close to
one up to temperature of about $170$ MeV, and slowly increases above that
temperature. Only at temperatures above $200$ MeV the deviations of this ratio from
unity are similar to the ones observed in strange and strange-charm sectors in the
transition region. This also suggests that significant thermal modifications of the
$J/\psi$ occur for $T\gtrsim200$ MeV. 

Finally, we comment on the sensitivity of spatial correlation functions to temporal
boundary conditions of the fermionic fields. At finite temperature, the temporal
boundary condition must be anti-periodic for the fermionic fields and all gauge field
configurations are generated by imposing such a boundary condition. 
On the other hand, on these
configurations one can measure observables also by imposing periodic temporal
boundary conditions for the fermion fields. Such a trick may provide some further
insight into the melting of charmonia states \cite{Mukherjee:2008tr,Karsch:2012na,Boyd:1994np}.
As discussed above, at asymptotically high temperatures the screening masses approach
twice the value of the lowest Matsubara frequency, that arises entirely from the
anti-periodic temporal boundary conditions for the fermions. This result reflects
that the two fermions (quarks) of the bosonic meson operator propagate independently
and are separately sensitive to the anti-periodic boundary conditions. This tells us
that if one measures the screening masses of the meson operators by artificially
imposing periodic temporal boundary conditions for the fermions, one expects
to find vanishing screening masses at very high temperatures.
Asymptotically, the quadratic difference of the screening masses calculated with
anti-periodic and periodic boundary conditions will approach $(2 \pi T)^2$
\cite{Mukherjee:2008tr,Karsch:2012na,Boyd:1994np}. More generally, by comparing screening masses
for mesonic observables calculated with both sets of boundary conditions we can probe
to what extent the corresponding correlators are influenced by the boundary
conditions, i.e. whether the fermionic substructure of the meson becomes visible and
influences the asymptotic behavior of spatial correlation functions. On the other
hand, if the two fermions constituting the meson operator remain as a well defined
bosonic bound state, then the corresponding screening mass should be insensitive to
the fermionic boundary conditions and it is expected to obtain identical screening
masses for both anti-periodic and periodic ones.  

In Fig.~\ref{fig:pb} we show the charmonium screening masses in the pseudo-scalar
($0^{-+}$) and vector ($1^{--}$) channels calculated using anti-periodic and periodic
boundary conditions. Both channels show similar temperature dependence: Already in
the crossover region the screening masses start to become sensitive to the boundary
conditions. However, there is little sensitivity to the boundary conditions for
$T\lesssim170$ MeV. Above that temperature we see clear sensitivity of the screening
masses to the boundary conditions, which becomes quite large for $T\gtrsim200$ MeV.
This may indicate the dissolution of the $\eta_c$ and $J/\psi$ states at these
temperatures.  However, to quantify the ``onset temperature'' for dissolution is
clearly not possible in this way. We also see sensitivity to the boundary conditions
in the scalar and axial-vector charmonium screening masses. But due to large errors
in the corresponding screening masses with the periodic boundary condition it is more
difficult to quantify this sensitivity.
      
\section{Conclusions}

We have studied spatial correlation functions at non-zero temperature for $s \bar s$, $s
\bar c$ and $c \bar c$ mesons to investigate their in-medium modifications.  
We have performed direct comparisons of the in-medium correlation functions with the
corresponding zero temperature ones, extracted screening masses and amplitudes from
large distance behaviors of the correlation functions and also investigated their
sensitivity to the temporal boundary conditions of the charm quark.  
In all cases, we have found that medium modifications of the spatial meson correlation functions set in the
crossover region. 
However we have also found that the amount of in-medium modifications in the spatial
correlators is different in different sectors and decreases with the heavy quark
content.  The $s \bar s$ and $s\bar c$ mesons are significantly effected by the
medium already at relatively low temperatures and possibly dissolve at temperature
close to the crossover temperature, $T_c=(154\pm9)$ MeV. For the $c \bar c$ mesons, S-wave charmonium states
($J/\psi$ and $\eta_c$) undergo very small medium modifications up to $T\sim1.1T_c$
($T\sim170$ MeV) and significant medium modifications are observed only for
$T\gtrsim1.3T_c$ ($T\gtrsim200$ MeV). We have also seen a clear
difference between the temperature dependence of the correlators corresponding to
S-wave charmonium and to P-wave charmonium ($\chi_{c0}$ and $\chi_{c1}$) states. The
spatial correlators corresponding to $\chi_{c0}$ and $\chi_{c1}$ states show sizable
medium modifications already in the crossover region. This is in line with the
sequential melting of charmonia states--- the larger, loosely bound P-wave states
dissociate at lower temperatures than the smaller, tightly bound S-wave charmonia.

Let us finally summurize the importance of our findings for the physics of heavy ion
collisions. The sequential charmonium melting is an essential ingredient for
most of phenomenological models that attempt to explain charmonium yield in heavy ion
collisions. Therefore our findings provide support for these models. The fact that 
open-charm mesons dissolve at temperatures close to the transition temperature
disfavors the models which try to explain the large energy loss of heavy quarks
in heavy ion collisions through the existence of heavy-light bound states in
quark gluon plasma \cite{Sharma:2009hn}. Finally the large medium modification 
of hidden strange meson correlators disfavors scenarios of separate freeze-out of strange
degrees of freedom in heavy ion collisions \cite{Chatterjee:2013yga}.

\section*{Acknowledgments} \label{ackn}

Numerical calculations were carried out on the USQCD Clusters at the Jefferson
Laboratory, USA, the NYBlue supercomputer at the Brookhaven National Laboratory, USA
and  in NERSC, USA. This work was partly supported by through the Contract No.
DE-AC02-98CH10886 with the U.S. Department of Energy and the Bundesministerium f\"ur
Bildung und Forschung under grant 05P12PBCTA and the EU Integrated Infrastructure
Initiative Hadron-Physics3.  Partial support for this work was also provided through
Scientific Discovery through Advanced Computing (SciDAC) program funded by U.S.
Department of Energy, Office of Science, Advanced Scientific Computing Research (and
Basic Energy Sciences/Biological and Environmental Research/High Energy
Physics/Fusion Energy Sciences/Nuclear Physics). 
The calculations reported in this paper have been performed using the public MILC code (MILC Collaboration: \url{http://www.physics.utah.edu/~detar/milc}).\vskip0.4truecm

\appendix

\section{Determination of charm quark mass}  \label{ap:mc}

\begin{figure}[!h]
\includegraphics[width=8.5cm]{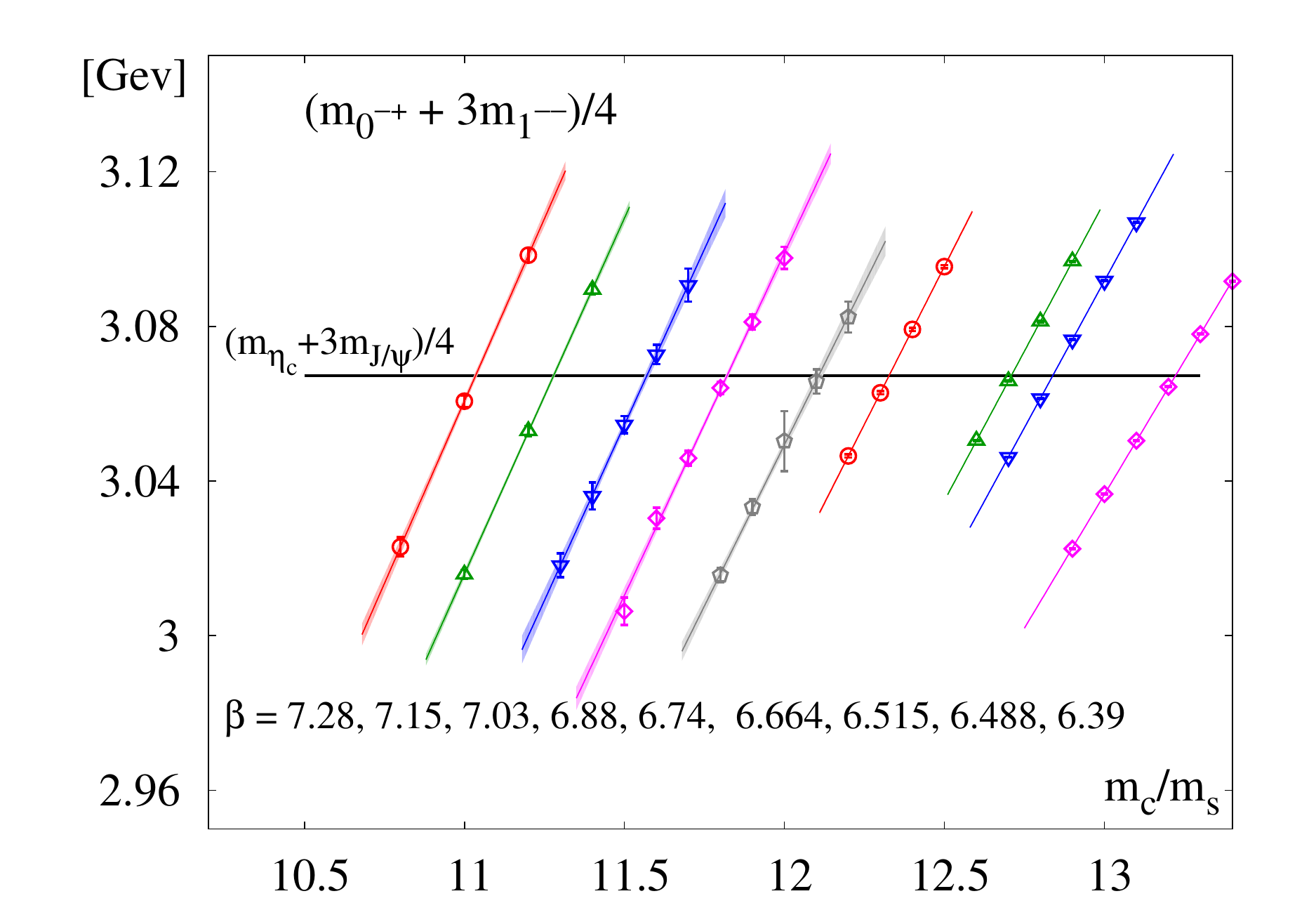}
\caption{Results for the spin averaged charmonium mass $(m_{0^{-+}} + 3m_{1^{--}})/4$
for several trial $m_c/m_s$ values of each $\beta$. The charm quark mass on LCP is
determined by matching those to the physical value $(m_{\eta_c}+3m_{J/\psi})/4$,
summarized in Tab.~\ref{tab:app_charm}.}
\label{fig:apmass}
\end{figure}

Here we determine the line of constant physics (LCP) for the charm quark mass.  To
estimate the charm quark mass, we calculate the masses of the negative parity ground
states, $J^{PC} = 0^{-+}$ and $1^{--}$ which correspond to $\eta_c$ and $J/\psi$
mesons, respectively, in a range of couplings $\beta=6.39$--7.28 with several trial
values for the ratio of the bare charm to strange quark masses, $m_c/m_s$, using
point sources.  We summarize the simulation parameters and lattice sizes in
Tab.~\ref{tab:app_charm}, and show results for the spin averaged charmonium mass
$(m_{0^{-+}} + 3m_{1^{--}})/4$ as a function of $m_c/m_s$ for each $\beta$ in
Fig.~\ref{fig:apmass}.  In the figure, we interpolate the masses linearly for each
$\beta$ and estimate the line of constant physics for the ratio $(m_c/m_s)_{\rm LCP}$
from the intersection with the physical value $(m_{\eta_c} + 3m_{J/\psi})/4$.  The
numerical results are also summarized in Tab.~\ref{tab:app_charm}.

\begin{figure}[!h]
\includegraphics[width=8.5cm]{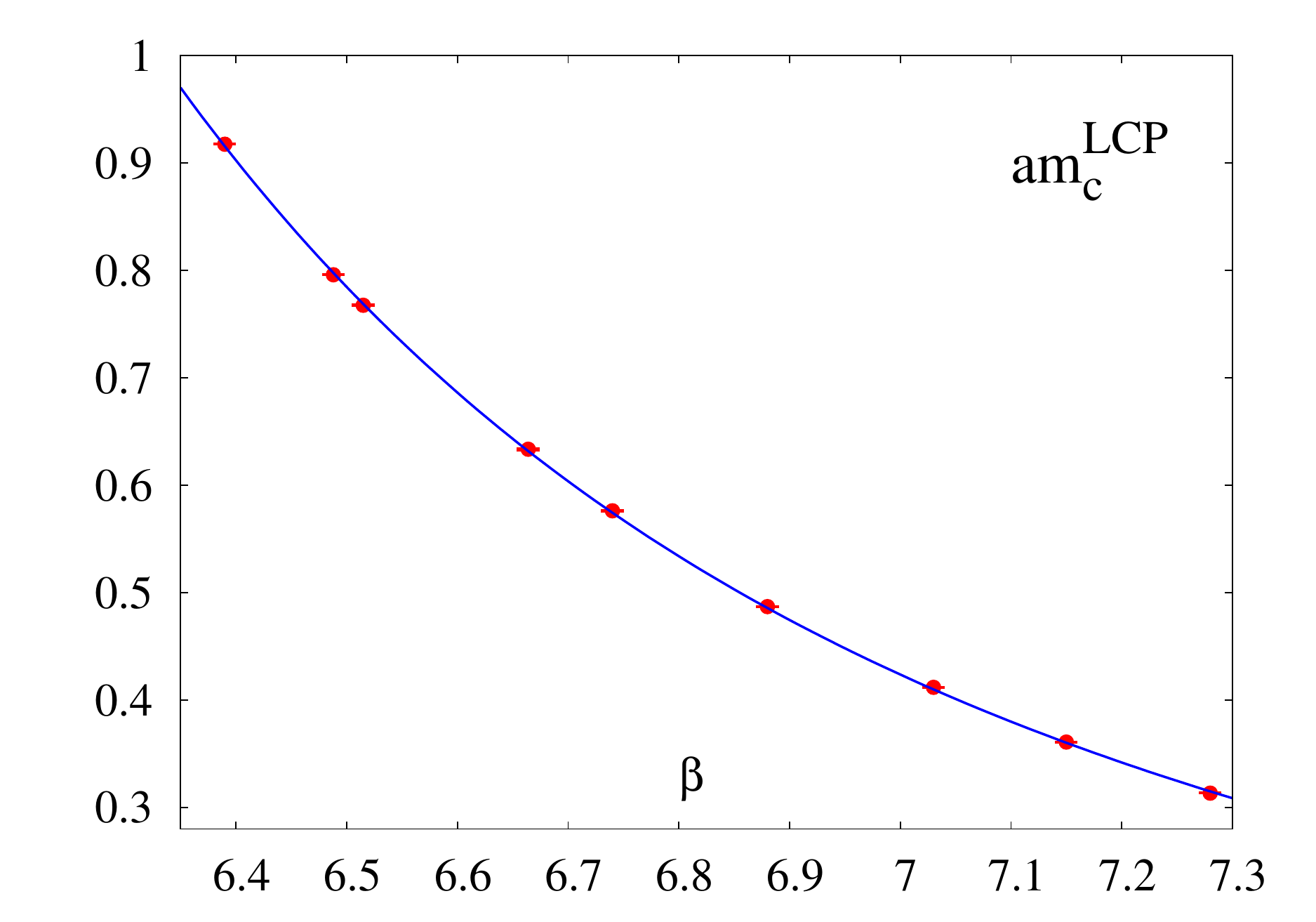}
\caption{Results for the charm quark mass on LCP as a function of $\beta$. The curve
shows the fit result (see text).}
\label{fig:apamc}
\end{figure}

From the ratio $(m_c/m_s)_{\rm LCP}$ we determine the charm quark mass on LCP shown
in Fig.~\ref{fig:apamc}.  We fit the data with the renormalization group inspired
form by Eq.~(\ref{eq:amc}) and obtain $c_0=61.0(1.7)$, $c_2=2.76(26)\times 10^5$ and
$d_2=3.3(3.7)\times10^2$. The fit result
is also shown in Fig.~\ref{fig:apamc} as a solid curve.

\begin{table}[h]
\caption{Gauge couplings ($\beta$), lattice sizes ($N_\sigma^3\times N_\tau$) and
strange quark mass ($m_s$) used to determine the charm quark mass.  Results of the
ratio of charm and strange quark masses $(m_c/m_s)_{\rm LCP}$ and the charm quark
mass $am_c^{\rm LCP}$ on the line of constant physics are also summarized.}
\begin{center}
\begin{tabular}{cclll}
\hline\hline
 \multicolumn{1}{c}{$\beta$} & 
 \multicolumn{1}{c}{$N_\sigma^3\times N_\tau$} & 
 \multicolumn{1}{c}{$am_s$} & 
 \multicolumn{1}{c}{$(m_c/m_s)_{\rm LCP}$} & 
 \multicolumn{1}{c}{$am_c^{\rm LCP}$} \\
\hline
6.390 & $32^4$ & 0.0694 & 13.2222(40)  & 0.91762(27)  \\
6.488 & $32^4$ & 0.0620 & 12.8386(33)  & 0.79600(21)  \\
6.515 & $32^4$ & 0.0604 & 12.7091(54)  & 0.76763(33)  \\
6.664 & $32^4$ & 0.0514 & 12.327(12)   & 0.63361(66)  \\
6.740 & $48^4$ & 0.0476 & 12.107(13)   & 0.57631(62)  \\
6.880 & $48^4$ & 0.0412 & 11.8208(53)  & 0.48702(22)  \\
7.030 & $48^4$ & 0.0356 & 11.5702(85)  & 0.41190(30)  \\
7.150 & $48^3\times64$ & 0.0320 & 11.2789(43)  & 0.36092(14)  \\
7.280 & $48^3\times64$ & 0.0284 & 11.0350(55)  & 0.31339(15)  \\
\hline\hline
\end{tabular}
\end{center}
\label{tab:app_charm}
\end{table}

\begin{figure}[!h]
\includegraphics[width=8.5cm]{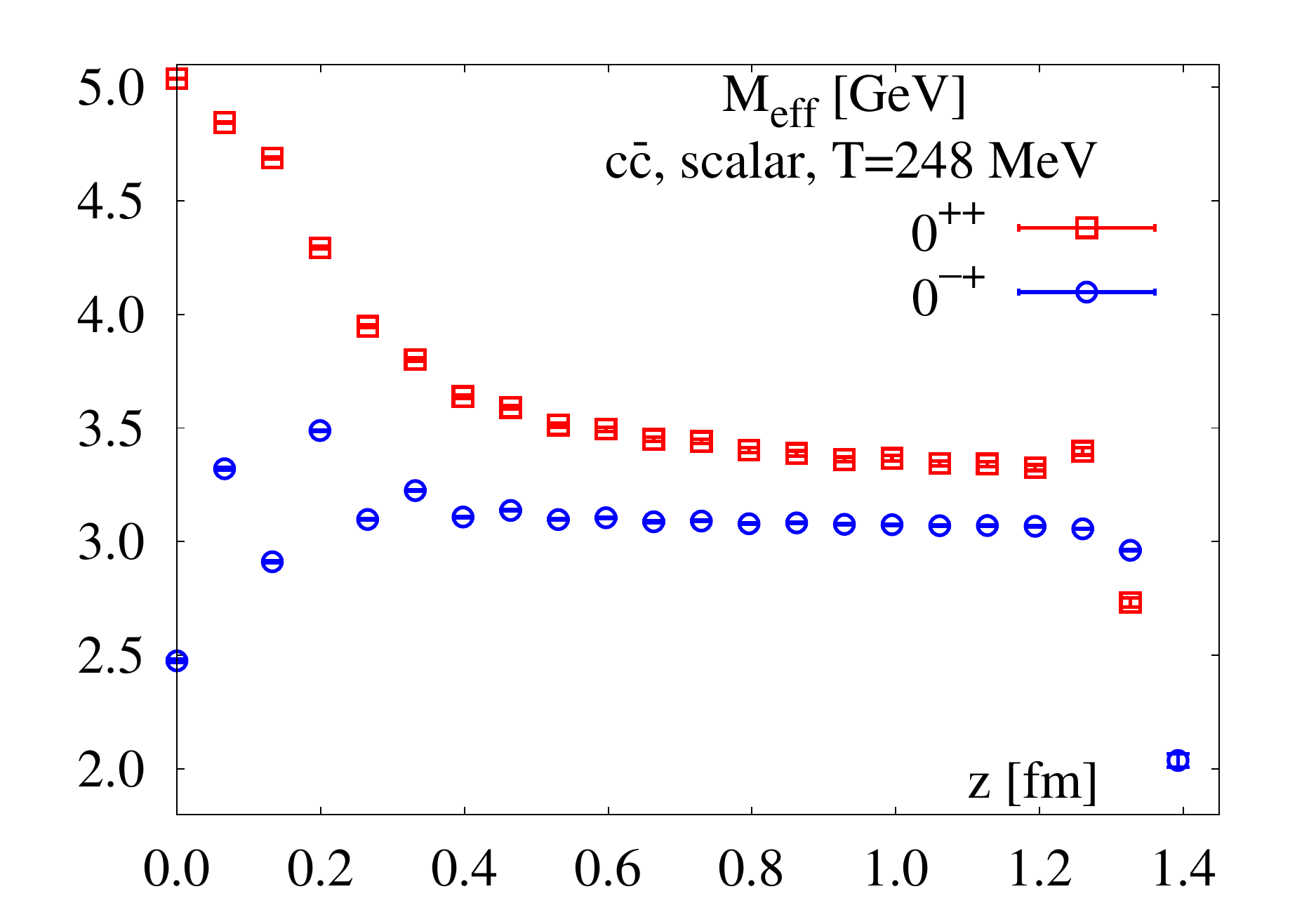}
\caption{Effective masses of the negative ($0^{-+}$) and positive ($0^{++}$) parity
states as functions of distance, for the $c\bar{c}$ scalar channel at $T=248$ MeV.}
\label{fig:apefmsks}
\end{figure}

\begin{figure}[!h]
\includegraphics[width=8.5cm]{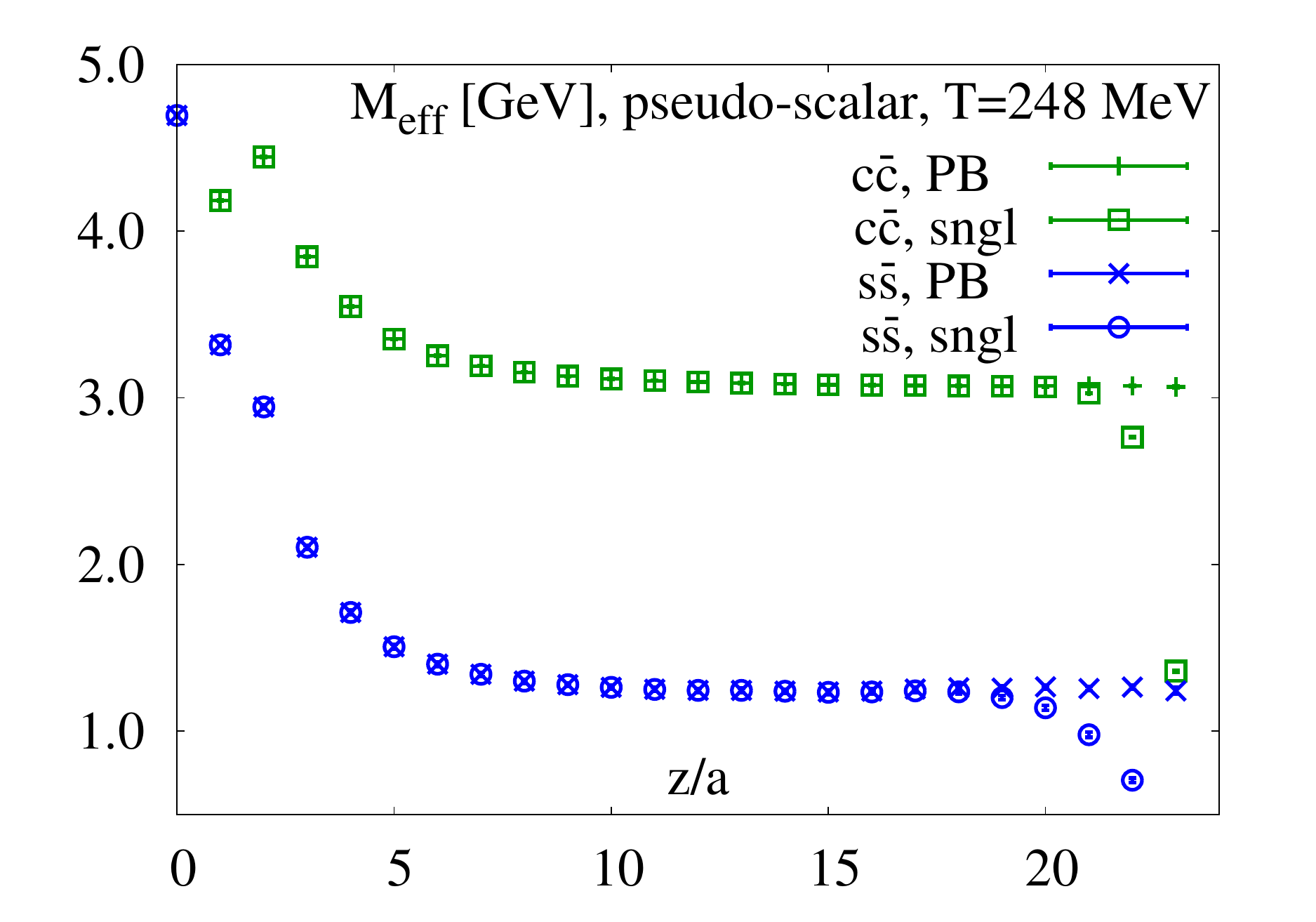}
\caption{The effective masses for the $s\bar{s}$ and $c\bar{c}$ pseudo-scalar channel
at $T=248$ MeV as functions of distance, obtained by including effects of periodicity
in the $z$-direction (PB) and also by using a single exponential decay ansatz (sngl).}
\label{fig:apefmsks_ps}
\end{figure}

\section{Negative and positive parity parts of staggered meson correlator from
effective masses} \label{ap:efms}

In this appendix we briefly discuss our procedure for identifying separate
contributions of the negative (non-oscillating) and positive (oscillating) parity
states in a given staggered meson correlation function. 

We assume that the staggered meson correlator can be described by a single negative
and a single positive parity meson states:
\begin{eqnarray}
G(z) = A_{NO}^2(z) e^{-M_-(z) z} -(-1)^z A_{O}^2(z) e^{-M_+(z) z}.
\label{eq:corr}
\end{eqnarray}
For simplicity, contributions arising from the periodicity of the lattice in the
$z$-direction have been neglected. Thus, the correlator is parametrized by four
quantities: $A_{NO}$, $A_O$ and $M_\pm$. 
Assuming that, for given value of $z$, these four
parameters vary little over at least four successive points in $z$, 
one can determine the parameters from values of the correlation function at these four points: $g_i\equiv
G(z+i)$ with $i=0,1,2,3$. With $x_\pm\equiv e^{-M_\pm}$ and the assumption $x_+\ne
x_-$, it is easy to see
\begin{eqnarray}
\frac{g_{i+2}-g_{i+1}x_-}{g_{i+1}-g_ix_-} = -x_+, \ \ \mathrm{for} \ \ i=0,1\;,
\end{eqnarray}
yielding a quadratic equation for $x_-$
\begin{eqnarray}
(g_2-g_1x_-)^2=(g_3-g_2x_-)(g_1-g_0x_-) \;.
\end{eqnarray} 
Similarly, one can obtain a quadratic equation for $x_+$
\begin{eqnarray}
(g_2+g_1x_+)^2=(g_3+g_2x_+)(g_1+g_0x_+)\; .
\end{eqnarray}
Both equations can be expressed in simple quadratic equations
\begin{eqnarray}
Ax_\pm^2\mp Bx_\pm +C=0\; ,
\end{eqnarray}
where $A=g_1^2-g_2g_0$, $B=g_3g_0-g_2g_1$ and $C=g_2^2-g_3g_1$.
Under a condition $A\cdot C<0$, definite solutions are given by
\begin{eqnarray}
x_\pm=\pm\frac{B}{2A} + \frac{\sqrt{B^2-4AC}}{2|A|}\; .
\end{eqnarray}
Once the local, effective masses $x_\pm$ are determined, it is possible to define two
separate effective correlators for the negative (non-oscillating) and positive
(oscillating) parity states of a staggered meson correlation function
\begin{subequations}
\begin{eqnarray}
G_{NO}(z) &\equiv& A_{NO}^2(z) e^{-M_-(z)z} =        \frac{g_1+g_0 x_+}{x_- + x_+} ,\\
G_{O}(z)  &\equiv& A_{O}^2(z) e^{-M_+(z)z}  = (-1)^z \frac{g_1-g_0 x_-}{x_- + x_+} .
\end{eqnarray}
\label{eq:effcorr}
\end{subequations}

As an illustrative example, in Fig.~\ref{fig:apefmsks} we show the local, effective
masses, $M_\pm=-\ln x_\pm$, obtained from the point source correlator of $c\bar{c}$
scalar channel at $T=248$ MeV as a function of $z$. At $T=0$ the positive parity
$0^{++}$ (negative parity $0^{-+}$) state corresponds to $\chi_{c0}$ (the different
taste of $\eta_c$). The effective masses for different parity states can be well
identified in this way, and plateaus appear at $z\ge0.5$ fm for the negative parity state
and at $z\ge0.8$ fm for the positive parity state. At large distances, $z\ge1.2$ fm,
the plateaus get worse due to the periodicity in the $z$-direction.

To estimate systematic errors that arise from neglecting the periodic terms in
Eq.~(\ref{eq:corr}) it is easier to look into the pseudo-scalar correlators in
$s\bar{s}$ and $c\bar{c}$ sectors, as they do not receive contribution from a
oscillating state. In this case, the correlation function is
parametrized as
\begin{eqnarray}
G(z) = A_{NO}^2(z)\left[ e^{-M_-(z)z} + e^{-M_-(z)(N_z-z)}\right]. 
\end{eqnarray}
The second term on the right-hand-side arises from  the periodic boundary conditions
in $z$-direction. If one neglects the periodicity, the effective mass is simply
obtained from the single exponential decay (sngl) form,
\begin{eqnarray}
M_-^\text{sngl}(z) = \ln \frac{G(z)}{G(z+1)}.
\end{eqnarray}
On the other hand, the effective mass including the effects of periodic boundary (PB)
can be determined from the numerical solution of the equation
\begin{eqnarray}
f(M_-^{\rm PB}) \equiv \frac{\cosh \left[M_-^{\rm PB}(z)
\left(z-\tfrac{N_z}{2}\right)\right]}{\cosh \left[M_-^{\rm PB}(z)
\left(z-\tfrac{N_z}{2}+1\right)\right]} - \frac{G(z)}{G(z+1)} = 0. \nonumber
\end{eqnarray}

In Fig.~\ref{fig:apefmsks_ps} we show results for the effective masses of the
$s\bar{s}$ and $c\bar{c}$ pseudo-scalar channels at $T=248$ MeV, obtained by
including the effects of periodic boundary (PB) as well as using the single
exponential decay (sngl) form. One finds that both effective masses converge at short
distances and reach a plateau at moderate values of the distance.  At large distances
$M_-^{\rm PB}$ stays on the plateau, whereas $M_-^{\rm sngl}$ deviates from the
plateau.  The deviation becomes visible at $z/a\sim18$ for $s\bar{s}$ and $\sim20$
for $c\bar{c}$. Thus, for our calculations, one can cleanly define the effective
correlators Eqs.~(\ref{eq:effcorr}) till $z/a<18$ for $s\bar{s}$ and till $z/a<20$
for $c\bar{c}$. We also performed a similar study for $s\bar{c}$ sector and find that
the deviation is visible only at $z/a\sim20$.  For positive parity states the
amplitudes are smaller than that for the negative parity states, and the affect of
the periodic boundary condition is slightly larger. Thus, to be conservative, we
apply our procedure for separating the contributions of the positive and negative
parity only up to $z/a < 15$.

\section{Summary of screening masses} \label{ap:cal}

The screening masses determined by fitting the corresponding correlators with the
corner-wall sources are summarized in Tabs.~\ref{tab:app2ss}, \ref{tab:app2sc} and
\ref{tab:app2cc} for $s\bar{s}$, $s\bar{c}$ and $c\bar{c}$ mesons, respectively.  The
1st and 2nd parentheses give the statistical and systematic errors.  The former is
estimated from a bootstrap analysis for each fit and the later gives the systematic
ambiguity arising from the variation of the fit range.

\begingroup
\begin{table*}[h!]
\caption{Numerical values of the screening masses, in GeV, for $s\bar{s}$ sector for
different temperatures. The 1st and 2nd parentheses indicate the statistical and
systematic errors, respectively.}
\begin{center}
\begin{tabular}{clllll}
\hline\hline
 \multicolumn{1}{c}{$T$ [MeV]} & 
 \multicolumn{1}{c}{$0^{-+}$} & 
 \multicolumn{1}{c}{$0^{++}$} & 
 \multicolumn{1}{c}{$1^{--}$} & 
 \multicolumn{1}{c}{$1^{++}$} \\
\hline
 138.2 & 0.69572(18)(35) &  1.0542(40)(96) &  1.0469(13)(29) &  1.3766(83)(50) \\
 143.3 &  0.70202(13)(4) &    1.041(2)(12) &   1.0565(8)(49) &    1.343(4)(16) \\
 149.0 & 0.70997(18)(39) &    1.016(2)(17) &  1.0618(11)(67) &    1.325(5)(23) \\
 153.5 & 0.71691(20)(33) &  1.0115(17)(97) &   1.0719(9)(47) &    1.300(3)(19) \\
 158.0 & 0.72777(21)(48) &  1.0010(15)(71) &   1.0852(9)(51) &    1.286(2)(19) \\
 164.3 & 0.74406(18)(62) &  0.9754(12)(64) &   1.0982(6)(55) &    1.273(2)(20) \\
 167.5 & 0.75755(34)(67) &  0.9715(15)(78) &  1.1146(14)(66) &    1.273(2)(16) \\
 170.8 & 0.76538(28)(27) &   0.9687(6)(51) &  1.1221(10)(67) &    1.268(1)(12) \\
 175.8 & 0.79145(40)(36) &  0.9690(11)(26) &   1.1475(9)(54) &  1.2742(20)(89) \\
 182.6 & 0.81736(51)(49) &   0.9717(8)(59) &   1.1673(9)(69) &    1.276(1)(16) \\
 189.6 &   0.8508(5)(11) &   0.9834(9)(32) &  1.2058(14)(69) &  1.2857(20)(85) \\
 196.9 &   0.8912(5)(11) &   1.0034(6)(39) &   1.2387(8)(50) &  1.3096(14)(85) \\
 210.2 &   0.9694(5)(11) &   1.0474(6)(33) &   1.3123(8)(32) &  1.3611(10)(61) \\
 220.2 &   1.0381(6)(12) &   1.0979(7)(23) &   1.3709(8)(31) &   1.4077(8)(51) \\
 247.9 & 1.23211(58)(73) &   1.2689(5)(10) &   1.5468(8)(41) &   1.5671(9)(49) \\
 297.5 &  1.5862(30)(17) &  1.5969(31)(13) &  1.8787(32)(35) &  1.8855(34)(40) \\
 371.9 &  2.1069(23)(35) &  2.1150(23)(50) &  2.3695(40)(28) &  2.3719(41)(29) \\
 495.8 &  2.9102(40)(55) &  2.9181(24)(37) &  3.1589(27)(47) &  3.1605(27)(51) \\
 743.7 &    4.512(6)(13) &  4.5233(55)(67) &  4.7677(45)(70) &  4.7674(45)(72) \\
\hline\hline
\end{tabular}
\end{center}
\label{tab:app2ss}
\end{table*}

\begin{table*}[h!]
\caption{Same as Tab.~\protect\ref{tab:app2ss}, but for $s\bar{c}$ sector.}
\begin{center}
\begin{tabular}{clllll}
\hline\hline
 \multicolumn{1}{c}{$T$ [MeV]} & 
 \multicolumn{1}{c}{$0^{-+}$} & 
 \multicolumn{1}{c}{$0^{++}$} & 
 \multicolumn{1}{c}{$1^{--}$} & 
 \multicolumn{1}{c}{$1^{++}$} \\
\hline
 138.2 & 1.97625(26)(97) &    2.326(3)(10) &   2.1176(5)(13) &    2.451(8)(14) \\
 143.3 & 1.97997(17)(81) &    2.293(2)(10) & 2.12105(47)(83) &  2.4255(53)(97) \\
 149.0 & 1.98459(28)(54) &  2.2789(25)(72) & 2.12494(46)(58) &    2.405(3)(11) \\
 153.5 & 1.98708(14)(67) &  2.2815(18)(42) & 2.12906(51)(86) &  2.3936(34)(89) \\
 158.0 & 1.99315(27)(88) &  2.2585(20)(45) & 2.13648(37)(64) &  2.3783(27)(71) \\
 164.3 &   2.0029(1)(10) &  2.2484(12)(40) & 2.14577(47)(64) &  2.3611(21)(67) \\
 167.5 & 2.00951(38)(46) &  2.2410(18)(38) &   2.1554(6)(12) &  2.3647(27)(63) \\
 170.8 &   2.0122(2)(11) &   2.2315(9)(33) & 2.15700(44)(50) &  2.3495(18)(47) \\
 175.8 & 2.02757(33)(98) &  2.2309(14)(19) & 2.17614(59)(69) &  2.3494(19)(59) \\
 182.6 &   2.0403(2)(14) &   2.2260(9)(29) & 2.18684(51)(96) &  2.3467(13)(60) \\
 189.6 &   2.0582(4)(11) &   2.2205(8)(12) & 2.20748(75)(87) &  2.3450(15)(32) \\
 196.9 &   2.0784(3)(20) &   2.2177(7)(18) &   2.2269(5)(20) &   2.3430(8)(35) \\
 210.2 &   2.1189(3)(18) &   2.2361(5)(16) &   2.2740(4)(23) &   2.3715(7)(11) \\
 220.2 &   2.1531(3)(19) &   2.2534(5)(14) &   2.3103(4)(15) &   2.3889(7)(19) \\
 247.9 &   2.2553(4)(31) &   2.3251(3)(21) &   2.4160(5)(17) &   2.4696(5)(10) \\
 297.5 &   2.4624(12)(9) &  2.5057(10)(20) &  2.6368(18)(24) &  2.6676(18)(35) \\
 371.9 &  2.7942(15)(35) &  2.8171(14)(38) &  2.9782(20)(25) &  2.9936(21)(24) \\
 495.8 &  3.4036(20)(54) &  3.4162(21)(50) &  3.6043(22)(67) &  3.6107(22)(66) \\
 743.7 &  4.8049(39)(92) &  4.8100(41)(93) &  5.0251(42)(97) &  5.0276(43)(98) \\
\hline\hline
\end{tabular}
\end{center}
\label{tab:app2sc}
\end{table*}

\begin{table*}[h!]
\caption{Same as Tab.~\protect\ref{tab:app2ss}, but for $c\bar{c}$ sector.}
\begin{center}
\begin{tabular}{clllll}
\hline\hline
 \multicolumn{1}{c}{$T$ [MeV]} & 
 \multicolumn{1}{c}{$0^{-+}$} & 
 \multicolumn{1}{c}{$0^{++}$} & 
 \multicolumn{1}{c}{$1^{--}$} & 
 \multicolumn{1}{c}{$1^{++}$} \\
\hline
138.2 &  2.97844(9)(27) &    3.406(9)(21) & 3.08523(20)(33) &   3.474(11)(23) \\
143.3 &  2.98062(7)(31) &    3.398(5)(13) & 3.08777(18)(34) &    3.487(9)(19) \\
149.0 &  2.98302(9)(50) &    3.392(5)(13) & 3.08999(16)(70) &    3.467(6)(17) \\
153.5 & 2.98150(10)(41) &    3.377(3)(14) & 3.08871(13)(67) &    3.452(7)(15) \\
158.0 &  2.98418(8)(60) &    3.366(4)(12) & 3.09144(16)(85) &    3.448(5)(16) \\
164.3 &  2.98748(7)(61) &    3.352(2)(10) & 3.09518(16)(84) &    3.419(4)(14) \\
167.5 & 2.98746(13)(51) &  3.3444(31)(66) & 3.09543(21)(40) &  3.4205(61)(91) \\
170.8 &  2.98641(8)(54) &  3.3405(21)(95) & 3.09487(13)(84) &    3.414(2)(11) \\
175.8 & 2.99113(12)(51) &  3.3468(28)(84) & 3.10050(21)(71) &  3.4108(38)(72) \\
182.6 & 2.99398(12)(73) &  3.3323(21)(75) & 3.10473(18)(87) &    3.394(2)(10) \\
189.6 & 2.99766(14)(63) &  3.3231(16)(84) & 3.10977(23)(85) &  3.3857(21)(75) \\
196.9 & 3.00230(14)(87) &  3.3031(14)(56) &   3.1154(2)(12) &  3.3657(18)(76) \\
210.2 &   3.0156(1)(10) &   3.2975(9)(40) &   3.1321(2)(12) &  3.3669(14)(49) \\
220.2 &   3.0284(1)(11) &   3.2914(8)(35) &   3.1474(2)(15) &   3.3644(9)(40) \\
247.9 &   3.0706(1)(14) &   3.3016(6)(29) &   3.1959(2)(20) &  3.3776(10)(37) \\
297.5 & 3.17340(62)(79) &  3.3635(14)(40) &   3.3126(7)(16) &  3.4564(20)(38) \\
371.9 &   3.3747(7)(31) &  3.5109(10)(35) &   3.5309(9)(31) &  3.6288(12)(38) \\
495.8 &  3.8299(14)(67) &  3.9185(19)(60) &  4.0168(18)(66) &  4.0725(27)(49) \\
743.7 &    5.044(5)(15) &    5.100(3)(10) &  5.2768(36)(82) &  5.2991(39)(83) \\
\hline\hline
\end{tabular}
\end{center}
\label{tab:app2cc}
\end{table*}
\endgroup

\bibliography{mscr}

\begin{thebibliography}{45}
\expandafter\ifx\csname natexlab\endcsname\relax\def\natexlab#1{#1}\fi
\expandafter\ifx\csname bibnamefont\endcsname\relax
  \def\bibnamefont#1{#1}\fi
\expandafter\ifx\csname bibfnamefont\endcsname\relax
  \def\bibfnamefont#1{#1}\fi
\expandafter\ifx\csname citenamefont\endcsname\relax
  \def\citenamefont#1{#1}\fi
\expandafter\ifx\csname url\endcsname\relax
  \def\url#1{\texttt{#1}}\fi
\expandafter\ifx\csname urlprefix\endcsname\relax\def\urlprefix{URL }\fi
\providecommand{\bibinfo}[2]{#2}
\providecommand{\eprint}[2][]{\url{#2}}

\bibitem[{\citenamefont{Bhattacharya et~al.}(2014)\citenamefont{Bhattacharya,
  Buchoff, Christ, Ding, Gupta et~al.}}]{Bhattacharya:2014ara}
\bibinfo{author}{\bibfnamefont{T.}~\bibnamefont{Bhattacharya}},
  \bibinfo{author}{\bibfnamefont{M.~I.} \bibnamefont{Buchoff}},
  \bibinfo{author}{\bibfnamefont{N.~H.} \bibnamefont{Christ}},
  \bibinfo{author}{\bibfnamefont{H.-T.} \bibnamefont{Ding}},
  \bibinfo{author}{\bibfnamefont{R.}~\bibnamefont{Gupta}},
  \bibnamefont{et~al.}, \bibinfo{journal}{Phys.Rev.Lett.}
  \textbf{\bibinfo{volume}{113}}, \bibinfo{pages}{082001}
  (\bibinfo{year}{2014}), \eprint{1402.5175}.

\bibitem[{\citenamefont{Bazavov
  et~al.}(2013{\natexlab{a}})\citenamefont{Bazavov, Ding, Hegde, Kaczmarek,
  Karsch et~al.}}]{Bazavov:2013dta}
\bibinfo{author}{\bibfnamefont{A.}~\bibnamefont{Bazavov}},
  \bibinfo{author}{\bibfnamefont{H.-T.} \bibnamefont{Ding}},
  \bibinfo{author}{\bibfnamefont{P.}~\bibnamefont{Hegde}},
  \bibinfo{author}{\bibfnamefont{O.}~\bibnamefont{Kaczmarek}},
  \bibinfo{author}{\bibfnamefont{F.}~\bibnamefont{Karsch}},
  \bibnamefont{et~al.}, \bibinfo{journal}{Phys.Rev.Lett.}
  \textbf{\bibinfo{volume}{111}}, \bibinfo{pages}{082301}
  (\bibinfo{year}{2013}{\natexlab{a}}), \eprint{1304.7220}.

\bibitem[{\citenamefont{Petreczky}(2012)}]{Petreczky:2012rq}
\bibinfo{author}{\bibfnamefont{P.}~\bibnamefont{Petreczky}},
  \bibinfo{journal}{J. Phys.} \textbf{\bibinfo{volume}{G39}},
  \bibinfo{pages}{093002} (\bibinfo{year}{2012}), \eprint{1203.5320}.

\bibitem[{\citenamefont{Philipsen}(2013)}]{Philipsen:2012nu}
\bibinfo{author}{\bibfnamefont{O.}~\bibnamefont{Philipsen}},
  \bibinfo{journal}{Prog. Part. Nucl. Phys.} \textbf{\bibinfo{volume}{70}},
  \bibinfo{pages}{55} (\bibinfo{year}{2013}), \eprint{1207.5999}.

\bibitem[{\citenamefont{Matsui and Satz}(1986)}]{Matsui:1986dk}
\bibinfo{author}{\bibfnamefont{T.}~\bibnamefont{Matsui}} \bibnamefont{and}
  \bibinfo{author}{\bibfnamefont{H.}~\bibnamefont{Satz}},
  \bibinfo{journal}{Phys. Lett.} \textbf{\bibinfo{volume}{B178}},
  \bibinfo{pages}{416} (\bibinfo{year}{1986}).

\bibitem[{\citenamefont{Sharma et~al.}(2009)\citenamefont{Sharma, Vitev, and
  Zhang}}]{Sharma:2009hn}
\bibinfo{author}{\bibfnamefont{R.}~\bibnamefont{Sharma}},
  \bibinfo{author}{\bibfnamefont{I.}~\bibnamefont{Vitev}}, \bibnamefont{and}
  \bibinfo{author}{\bibfnamefont{B.-W.} \bibnamefont{Zhang}},
  \bibinfo{journal}{Phys. Rev.} \textbf{\bibinfo{volume}{C80}},
  \bibinfo{pages}{054902} (\bibinfo{year}{2009}), \eprint{0904.0032}.

\bibitem[{\citenamefont{Bazavov
  et~al.}(2014{\natexlab{a}})\citenamefont{Bazavov, Ding, Hegde, Kaczmarek,
  Karsch et~al.}}]{Bazavov:2014yba}
\bibinfo{author}{\bibfnamefont{A.}~\bibnamefont{Bazavov}},
  \bibinfo{author}{\bibfnamefont{H.-T.} \bibnamefont{Ding}},
  \bibinfo{author}{\bibfnamefont{P.}~\bibnamefont{Hegde}},
  \bibinfo{author}{\bibfnamefont{O.}~\bibnamefont{Kaczmarek}},
  \bibinfo{author}{\bibfnamefont{F.}~\bibnamefont{Karsch}},
  \bibnamefont{et~al.}, \bibinfo{journal}{Phys.Lett.}
  \textbf{\bibinfo{volume}{B737}}, \bibinfo{pages}{210}
  (\bibinfo{year}{2014}{\natexlab{a}}), \eprint{1404.4043}.

\bibitem[{\citenamefont{Detar and Kogut}(1987{\natexlab{a}})}]{DeTar:1987ar}
\bibinfo{author}{\bibfnamefont{C.~E.} \bibnamefont{Detar}} \bibnamefont{and}
  \bibinfo{author}{\bibfnamefont{J.~B.} \bibnamefont{Kogut}},
  \bibinfo{journal}{Phys. Rev. Lett.} \textbf{\bibinfo{volume}{59}},
  \bibinfo{pages}{399} (\bibinfo{year}{1987}{\natexlab{a}}).

\bibitem[{\citenamefont{Detar and Kogut}(1987{\natexlab{b}})}]{DeTar:1987xb}
\bibinfo{author}{\bibfnamefont{C.~E.} \bibnamefont{Detar}} \bibnamefont{and}
  \bibinfo{author}{\bibfnamefont{J.~B.} \bibnamefont{Kogut}},
  \bibinfo{journal}{Phys. Rev.} \textbf{\bibinfo{volume}{D36}},
  \bibinfo{pages}{2828} (\bibinfo{year}{1987}{\natexlab{b}}).

\bibitem[{\citenamefont{Laine and Vepsalainen}(2004)}]{Laine:2003bd}
\bibinfo{author}{\bibfnamefont{M.}~\bibnamefont{Laine}} \bibnamefont{and}
  \bibinfo{author}{\bibfnamefont{M.}~\bibnamefont{Vepsalainen}},
  \bibinfo{journal}{JHEP} \textbf{\bibinfo{volume}{0402}}, \bibinfo{pages}{004}
  (\bibinfo{year}{2004}), \eprint{hep-ph/0311268}.

\bibitem[{\citenamefont{Brandt et~al.}(2014)\citenamefont{Brandt, Francis,
  Laine, and Meyer}}]{Brandt:2014uda}
\bibinfo{author}{\bibfnamefont{B.}~\bibnamefont{Brandt}},
  \bibinfo{author}{\bibfnamefont{A.}~\bibnamefont{Francis}},
  \bibinfo{author}{\bibfnamefont{M.}~\bibnamefont{Laine}}, \bibnamefont{and}
  \bibinfo{author}{\bibfnamefont{H.}~\bibnamefont{Meyer}},
  \bibinfo{journal}{JHEP} \textbf{\bibinfo{volume}{1405}}, \bibinfo{pages}{117}
  (\bibinfo{year}{2014}), \eprint{1404.2404}.

\bibitem[{\citenamefont{Asakawa et~al.}(2001)\citenamefont{Asakawa, Hatsuda,
  and Nakahara}}]{Asakawa:2000tr}
\bibinfo{author}{\bibfnamefont{M.}~\bibnamefont{Asakawa}},
  \bibinfo{author}{\bibfnamefont{T.}~\bibnamefont{Hatsuda}}, \bibnamefont{and}
  \bibinfo{author}{\bibfnamefont{Y.}~\bibnamefont{Nakahara}},
  \bibinfo{journal}{Prog. Part. Nucl. Phys.} \textbf{\bibinfo{volume}{46}},
  \bibinfo{pages}{459} (\bibinfo{year}{2001}), \eprint{hep-lat/0011040}.

\bibitem[{\citenamefont{Wetzorke et~al.}(2002)\citenamefont{Wetzorke, Karsch,
  Laermann, Petreczky, and Stickan}}]{Wetzorke:2001dk}
\bibinfo{author}{\bibfnamefont{I.}~\bibnamefont{Wetzorke}},
  \bibinfo{author}{\bibfnamefont{F.}~\bibnamefont{Karsch}},
  \bibinfo{author}{\bibfnamefont{E.}~\bibnamefont{Laermann}},
  \bibinfo{author}{\bibfnamefont{P.}~\bibnamefont{Petreczky}},
  \bibnamefont{and} \bibinfo{author}{\bibfnamefont{S.}~\bibnamefont{Stickan}},
  \bibinfo{journal}{Nucl. Phys. Proc. Suppl.} \textbf{\bibinfo{volume}{106}},
  \bibinfo{pages}{510} (\bibinfo{year}{2002}), \eprint{hep-lat/0110132}.

\bibitem[{\citenamefont{Asakawa et~al.}(2003)\citenamefont{Asakawa, Hatsuda,
  and Nakahara}}]{Asakawa:2002xj}
\bibinfo{author}{\bibfnamefont{M.}~\bibnamefont{Asakawa}},
  \bibinfo{author}{\bibfnamefont{T.}~\bibnamefont{Hatsuda}}, \bibnamefont{and}
  \bibinfo{author}{\bibfnamefont{Y.}~\bibnamefont{Nakahara}},
  \bibinfo{journal}{Nucl. Phys.} \textbf{\bibinfo{volume}{A715}},
  \bibinfo{pages}{863} (\bibinfo{year}{2003}), \eprint{hep-lat/0208059}.

\bibitem[{\citenamefont{Asakawa and Hatsuda}(2004)}]{Asakawa:2003re}
\bibinfo{author}{\bibfnamefont{M.}~\bibnamefont{Asakawa}} \bibnamefont{and}
  \bibinfo{author}{\bibfnamefont{T.}~\bibnamefont{Hatsuda}},
  \bibinfo{journal}{Phys. Rev. Lett.} \textbf{\bibinfo{volume}{92}},
  \bibinfo{pages}{012001} (\bibinfo{year}{2004}), \eprint{hep-lat/0308034}.

\bibitem[{\citenamefont{Datta et~al.}(2004)\citenamefont{Datta, Karsch,
  Petreczky, and Wetzorke}}]{Datta:2003ww}
\bibinfo{author}{\bibfnamefont{S.}~\bibnamefont{Datta}},
  \bibinfo{author}{\bibfnamefont{F.}~\bibnamefont{Karsch}},
  \bibinfo{author}{\bibfnamefont{P.}~\bibnamefont{Petreczky}},
  \bibnamefont{and} \bibinfo{author}{\bibfnamefont{I.}~\bibnamefont{Wetzorke}},
  \bibinfo{journal}{Phys. Rev.} \textbf{\bibinfo{volume}{D69}},
  \bibinfo{pages}{094507} (\bibinfo{year}{2004}), \eprint{hep-lat/0312037}.

\bibitem[{\citenamefont{Jakovac et~al.}(2007)\citenamefont{Jakovac, Petreczky,
  Petrov, and Velytsky}}]{Jakovac:2006sf}
\bibinfo{author}{\bibfnamefont{A.}~\bibnamefont{Jakovac}},
  \bibinfo{author}{\bibfnamefont{P.}~\bibnamefont{Petreczky}},
  \bibinfo{author}{\bibfnamefont{K.}~\bibnamefont{Petrov}}, \bibnamefont{and}
  \bibinfo{author}{\bibfnamefont{A.}~\bibnamefont{Velytsky}},
  \bibinfo{journal}{Phys. Rev.} \textbf{\bibinfo{volume}{D75}},
  \bibinfo{pages}{014506} (\bibinfo{year}{2007}), \eprint{hep-lat/0611017}.

\bibitem[{\citenamefont{Aarts et~al.}(2007)\citenamefont{Aarts, Allton, Oktay,
  Peardon, and Skullerud}}]{Aarts:2007pk}
\bibinfo{author}{\bibfnamefont{G.}~\bibnamefont{Aarts}},
  \bibinfo{author}{\bibfnamefont{C.}~\bibnamefont{Allton}},
  \bibinfo{author}{\bibfnamefont{M.~B.} \bibnamefont{Oktay}},
  \bibinfo{author}{\bibfnamefont{M.}~\bibnamefont{Peardon}}, \bibnamefont{and}
  \bibinfo{author}{\bibfnamefont{J.-I.} \bibnamefont{Skullerud}},
  \bibinfo{journal}{Phys. Rev.} \textbf{\bibinfo{volume}{D76}},
  \bibinfo{pages}{094513} (\bibinfo{year}{2007}), \eprint{0705.2198}.

\bibitem[{\citenamefont{Ding et~al.}(2012)\citenamefont{Ding, Francis,
  Kaczmarek, Karsch, Satz et~al.}}]{Ding:2012sp}
\bibinfo{author}{\bibfnamefont{H.-T.} \bibnamefont{Ding}},
  \bibinfo{author}{\bibfnamefont{A.}~\bibnamefont{Francis}},
  \bibinfo{author}{\bibfnamefont{O.}~\bibnamefont{Kaczmarek}},
  \bibinfo{author}{\bibfnamefont{F.}~\bibnamefont{Karsch}},
  \bibinfo{author}{\bibfnamefont{H.}~\bibnamefont{Satz}}, \bibnamefont{et~al.},
  \bibinfo{journal}{Phys. Rev.} \textbf{\bibinfo{volume}{D86}},
  \bibinfo{pages}{014509} (\bibinfo{year}{2012}), \eprint{1204.4945}.

\bibitem[{\citenamefont{Petreczky}(2009)}]{Petreczky:2008px}
\bibinfo{author}{\bibfnamefont{P.}~\bibnamefont{Petreczky}},
  \bibinfo{journal}{Eur. Phys. J.} \textbf{\bibinfo{volume}{C62}},
  \bibinfo{pages}{85} (\bibinfo{year}{2009}), \eprint{0810.0258}.

\bibitem[{\citenamefont{Florkowski and Friman}(1994)}]{Florkowski:1993bq}
\bibinfo{author}{\bibfnamefont{W.}~\bibnamefont{Florkowski}} \bibnamefont{and}
  \bibinfo{author}{\bibfnamefont{B.~L.} \bibnamefont{Friman}},
  \bibinfo{journal}{Z.Phys.} \textbf{\bibinfo{volume}{A347}},
  \bibinfo{pages}{271} (\bibinfo{year}{1994}).

\bibitem[{\citenamefont{Kaczmarek et~al.}(2014)\citenamefont{Kaczmarek,
  Laermann, and Müller}}]{Kaczmarek:2013kva}
\bibinfo{author}{\bibfnamefont{O.}~\bibnamefont{Kaczmarek}},
  \bibinfo{author}{\bibfnamefont{E.}~\bibnamefont{Laermann}}, \bibnamefont{and}
  \bibinfo{author}{\bibfnamefont{M.}~\bibnamefont{Müller}},
  \bibinfo{journal}{PoS} \textbf{\bibinfo{volume}{LATTICE2013}},
  \bibinfo{pages}{150} (\bibinfo{year}{2014}), \eprint{1311.3889}.

\bibitem[{\citenamefont{Gavai et~al.}(2008)\citenamefont{Gavai, Gupta, and
  Lacaze}}]{Gavai:2008yv}
\bibinfo{author}{\bibfnamefont{R.}~\bibnamefont{Gavai}},
  \bibinfo{author}{\bibfnamefont{S.}~\bibnamefont{Gupta}}, \bibnamefont{and}
  \bibinfo{author}{\bibfnamefont{R.}~\bibnamefont{Lacaze}},
  \bibinfo{journal}{Phys.Rev.} \textbf{\bibinfo{volume}{D78}},
  \bibinfo{pages}{014502} (\bibinfo{year}{2008}), \eprint{0803.1368}.

\bibitem[{\citenamefont{Banerjee et~al.}(2011)\citenamefont{Banerjee, Gavai,
  and Gupta}}]{Banerjee:2011yd}
\bibinfo{author}{\bibfnamefont{D.}~\bibnamefont{Banerjee}},
  \bibinfo{author}{\bibfnamefont{R.~V.} \bibnamefont{Gavai}}, \bibnamefont{and}
  \bibinfo{author}{\bibfnamefont{S.}~\bibnamefont{Gupta}},
  \bibinfo{journal}{Phys.Rev.} \textbf{\bibinfo{volume}{D83}},
  \bibinfo{pages}{074510} (\bibinfo{year}{2011}), \eprint{1102.4465}.

\bibitem[{\citenamefont{Iida et~al.}(2010)\citenamefont{Iida, Maezawa, and
  Yazaki}}]{Iida:2010jz}
\bibinfo{author}{\bibfnamefont{H.}~\bibnamefont{Iida}},
  \bibinfo{author}{\bibfnamefont{Y.}~\bibnamefont{Maezawa}}, \bibnamefont{and}
  \bibinfo{author}{\bibfnamefont{K.}~\bibnamefont{Yazaki}},
  \bibinfo{journal}{PoS} \textbf{\bibinfo{volume}{LATTICE2010}},
  \bibinfo{pages}{189} (\bibinfo{year}{2010}), \eprint{1012.2044}.

\bibitem[{\citenamefont{Cheng et~al.}(2011)\citenamefont{Cheng, Datta, Francis,
  van~der Heide, Jung et~al.}}]{Cheng:2010fe}
\bibinfo{author}{\bibfnamefont{M.}~\bibnamefont{Cheng}},
  \bibinfo{author}{\bibfnamefont{S.}~\bibnamefont{Datta}},
  \bibinfo{author}{\bibfnamefont{A.}~\bibnamefont{Francis}},
  \bibinfo{author}{\bibfnamefont{J.}~\bibnamefont{van~der Heide}},
  \bibinfo{author}{\bibfnamefont{C.}~\bibnamefont{Jung}}, \bibnamefont{et~al.},
  \bibinfo{journal}{Eur. Phys. J.} \textbf{\bibinfo{volume}{C71}},
  \bibinfo{pages}{1564} (\bibinfo{year}{2011}), \eprint{1010.1216}.

\bibitem[{\citenamefont{Laermann and Pucci}(2012)}]{Laermann:2012sr}
\bibinfo{author}{\bibfnamefont{E.}~\bibnamefont{Laermann}} \bibnamefont{and}
  \bibinfo{author}{\bibfnamefont{F.}~\bibnamefont{Pucci}},
  \bibinfo{journal}{Eur.Phys.J.} \textbf{\bibinfo{volume}{C72}},
  \bibinfo{pages}{2200} (\bibinfo{year}{2012}), \eprint{1207.6615}.

\bibitem[{\citenamefont{Karsch et~al.}(2012)\citenamefont{Karsch, Laermann,
  Mukherjee, and Petreczky}}]{Karsch:2012na}
\bibinfo{author}{\bibfnamefont{F.}~\bibnamefont{Karsch}},
  \bibinfo{author}{\bibfnamefont{E.}~\bibnamefont{Laermann}},
  \bibinfo{author}{\bibfnamefont{S.}~\bibnamefont{Mukherjee}},
  \bibnamefont{and}
  \bibinfo{author}{\bibfnamefont{P.}~\bibnamefont{Petreczky}},
  \bibinfo{journal}{Phys. Rev.} \textbf{\bibinfo{volume}{D85}},
  \bibinfo{pages}{114501} (\bibinfo{year}{2012}), \eprint{1203.3770}.

\bibitem[{\citenamefont{Follana et~al.}(2007)}]{Follana:2006rc}
\bibinfo{author}{\bibfnamefont{E.}~\bibnamefont{Follana}} \bibnamefont{et~al.}
  (\bibinfo{collaboration}{HPQCD Collaboration, UKQCD Collaboration}),
  \bibinfo{journal}{Phys. Rev.} \textbf{\bibinfo{volume}{D75}},
  \bibinfo{pages}{054502} (\bibinfo{year}{2007}), \eprint{hep-lat/0610092}.

\bibitem[{\citenamefont{Bazavov
  et~al.}(2012{\natexlab{a}})\citenamefont{Bazavov, Bhattacharya, Cheng, DeTar,
  Ding et~al.}}]{Bazavov:2011nk}
\bibinfo{author}{\bibfnamefont{A.}~\bibnamefont{Bazavov}},
  \bibinfo{author}{\bibfnamefont{T.}~\bibnamefont{Bhattacharya}},
  \bibinfo{author}{\bibfnamefont{M.}~\bibnamefont{Cheng}},
  \bibinfo{author}{\bibfnamefont{C.}~\bibnamefont{DeTar}},
  \bibinfo{author}{\bibfnamefont{H.-T.} \bibnamefont{Ding}},
  \bibnamefont{et~al.}, \bibinfo{journal}{Phys. Rev.}
  \textbf{\bibinfo{volume}{D85}}, \bibinfo{pages}{054503}
  (\bibinfo{year}{2012}{\natexlab{a}}), \eprint{1111.1710}.

\bibitem[{\citenamefont{Davies et~al.}(2012)\citenamefont{Davies, Donald,
  Dowdall, Koponen, Follana et~al.}}]{Davies:2013ju}
\bibinfo{author}{\bibfnamefont{C.}~\bibnamefont{Davies}},
  \bibinfo{author}{\bibfnamefont{G.}~\bibnamefont{Donald}},
  \bibinfo{author}{\bibfnamefont{R.}~\bibnamefont{Dowdall}},
  \bibinfo{author}{\bibfnamefont{J.}~\bibnamefont{Koponen}},
  \bibinfo{author}{\bibfnamefont{E.}~\bibnamefont{Follana}},
  \bibnamefont{et~al.}, \bibinfo{journal}{PoS}
  \textbf{\bibinfo{volume}{ConfinementX}}, \bibinfo{pages}{288}
  (\bibinfo{year}{2012}), \eprint{1301.7203}.

\bibitem[{\citenamefont{Bazavov et~al.}(2013{\natexlab{b}})}]{Bazavov:2013nfa}
\bibinfo{author}{\bibfnamefont{A.}~\bibnamefont{Bazavov}} \bibnamefont{et~al.}
  (\bibinfo{collaboration}{Fermilab Lattice and MILC Collaborations})
  (\bibinfo{year}{2013}{\natexlab{b}}), \eprint{1312.0149}.

\bibitem[{\citenamefont{Maezawa et~al.}(2013)\citenamefont{Maezawa, Bazavov,
  Karsch, Petreczky, and Mukherjee}}]{Maezawa:2013nxa}
\bibinfo{author}{\bibfnamefont{Y.}~\bibnamefont{Maezawa}},
  \bibinfo{author}{\bibfnamefont{A.}~\bibnamefont{Bazavov}},
  \bibinfo{author}{\bibfnamefont{F.}~\bibnamefont{Karsch}},
  \bibinfo{author}{\bibfnamefont{P.}~\bibnamefont{Petreczky}},
  \bibnamefont{and} \bibinfo{author}{\bibfnamefont{S.}~\bibnamefont{Mukherjee}}
  (\bibinfo{year}{2013}), \eprint{1312.4375}.

\bibitem[{\citenamefont{Bazavov
  et~al.}(2014{\natexlab{b}})\citenamefont{Bazavov, Karsch, Maezawa, Mukherjee,
  and Petreczky}}]{Bazavov:2014uda}
\bibinfo{author}{\bibfnamefont{A.}~\bibnamefont{Bazavov}},
  \bibinfo{author}{\bibfnamefont{F.}~\bibnamefont{Karsch}},
  \bibinfo{author}{\bibfnamefont{Y.}~\bibnamefont{Maezawa}},
  \bibinfo{author}{\bibfnamefont{S.}~\bibnamefont{Mukherjee}},
  \bibnamefont{and}
  \bibinfo{author}{\bibfnamefont{P.}~\bibnamefont{Petreczky}},
  \bibinfo{journal}{J.Phys.Conf.Ser.} \textbf{\bibinfo{volume}{535}},
  \bibinfo{pages}{012031} (\bibinfo{year}{2014}{\natexlab{b}}).

\bibitem[{\citenamefont{Altmeyer et~al.}(1993)}]{Altmeyer:1992dd}
\bibinfo{author}{\bibfnamefont{R.}~\bibnamefont{Altmeyer}} \bibnamefont{et~al.}
  (\bibinfo{collaboration}{MT(c) collaboration}), \bibinfo{journal}{Nucl.
  Phys.} \textbf{\bibinfo{volume}{B389}}, \bibinfo{pages}{445}
  (\bibinfo{year}{1993}).

\bibitem[{\citenamefont{Lepage et~al.}(2002)\citenamefont{Lepage, Clark,
  Davies, Hornbostel, Mackenzie et~al.}}]{Lepage:2001ym}
\bibinfo{author}{\bibfnamefont{G.}~\bibnamefont{Lepage}},
  \bibinfo{author}{\bibfnamefont{B.}~\bibnamefont{Clark}},
  \bibinfo{author}{\bibfnamefont{C.}~\bibnamefont{Davies}},
  \bibinfo{author}{\bibfnamefont{K.}~\bibnamefont{Hornbostel}},
  \bibinfo{author}{\bibfnamefont{P.}~\bibnamefont{Mackenzie}},
  \bibnamefont{et~al.}, \bibinfo{journal}{Nucl. Phys. Proc. Suppl.}
  \textbf{\bibinfo{volume}{106}}, \bibinfo{pages}{12} (\bibinfo{year}{2002}),
  \eprint{hep-lat/0110175}.

\bibitem[{\citenamefont{Levkova and DeTar}(2011)}]{Levkova:2010ft}
\bibinfo{author}{\bibfnamefont{L.}~\bibnamefont{Levkova}} \bibnamefont{and}
  \bibinfo{author}{\bibfnamefont{C.}~\bibnamefont{DeTar}},
  \bibinfo{journal}{Phys. Rev.} \textbf{\bibinfo{volume}{D83}},
  \bibinfo{pages}{074504} (\bibinfo{year}{2011}), \eprint{1012.1837}.

\bibitem[{\citenamefont{Bazavov et~al.}(2012{\natexlab{b}})}]{Bazavov:2012jq}
\bibinfo{author}{\bibfnamefont{A.}~\bibnamefont{Bazavov}} \bibnamefont{et~al.}
  (\bibinfo{collaboration}{HotQCD Collaboration}), \bibinfo{journal}{Phys.
  Rev.} \textbf{\bibinfo{volume}{D86}}, \bibinfo{pages}{034509}
  (\bibinfo{year}{2012}{\natexlab{b}}), \eprint{1203.0784}.

\bibitem[{\citenamefont{Bazavov et~al.}(2014{\natexlab{c}})}]{Bazavov:2014pvz}
\bibinfo{author}{\bibfnamefont{A.}~\bibnamefont{Bazavov}} \bibnamefont{et~al.}
  (\bibinfo{collaboration}{HotQCD Collaboration}), \bibinfo{journal}{Phys.Rev.}
  \textbf{\bibinfo{volume}{D90}}, \bibinfo{pages}{094503}
  (\bibinfo{year}{2014}{\natexlab{c}}), \eprint{1407.6387}.

\bibitem[{\citenamefont{Beringer et~al.}(2012)}]{Beringer:1900zz}
\bibinfo{author}{\bibfnamefont{J.}~\bibnamefont{Beringer}} \bibnamefont{et~al.}
  (\bibinfo{collaboration}{Particle Data Group}), \bibinfo{journal}{Phys. Rev.}
  \textbf{\bibinfo{volume}{D86}}, \bibinfo{pages}{010001}
  (\bibinfo{year}{2012}).

\bibitem[{\citenamefont{Davies et~al.}(2010)\citenamefont{Davies, McNeile,
  Wong, Follana, Horgan et~al.}}]{Davies:2009ih}
\bibinfo{author}{\bibfnamefont{C.}~\bibnamefont{Davies}},
  \bibinfo{author}{\bibfnamefont{C.}~\bibnamefont{McNeile}},
  \bibinfo{author}{\bibfnamefont{K.}~\bibnamefont{Wong}},
  \bibinfo{author}{\bibfnamefont{E.}~\bibnamefont{Follana}},
  \bibinfo{author}{\bibfnamefont{R.}~\bibnamefont{Horgan}},
  \bibnamefont{et~al.}, \bibinfo{journal}{Phys. Rev. Lett.}
  \textbf{\bibinfo{volume}{104}}, \bibinfo{pages}{132003}
  (\bibinfo{year}{2010}), \eprint{0910.3102}.

\bibitem[{\citenamefont{Wang et~al.}(2013)\citenamefont{Wang, Liu, Chang,
  Roberts, and Schmidt}}]{Wang:2013wk}
\bibinfo{author}{\bibfnamefont{K.-l.} \bibnamefont{Wang}},
  \bibinfo{author}{\bibfnamefont{Y.-x.} \bibnamefont{Liu}},
  \bibinfo{author}{\bibfnamefont{L.}~\bibnamefont{Chang}},
  \bibinfo{author}{\bibfnamefont{C.~D.} \bibnamefont{Roberts}},
  \bibnamefont{and} \bibinfo{author}{\bibfnamefont{S.~M.}
  \bibnamefont{Schmidt}}, \bibinfo{journal}{Phys. Rev.}
  \textbf{\bibinfo{volume}{D87}}, \bibinfo{pages}{074038}
  (\bibinfo{year}{2013}), \eprint{1301.6762}.

\bibitem[{\citenamefont{Mukherjee}(2009)}]{Mukherjee:2008tr}
\bibinfo{author}{\bibfnamefont{S.}~\bibnamefont{Mukherjee}},
  \bibinfo{journal}{Nucl.Phys.} \textbf{\bibinfo{volume}{A820}},
  \bibinfo{pages}{283C} (\bibinfo{year}{2009}), \eprint{0810.2906}.

\bibitem[{\citenamefont{Boyd et~al.}(1994)\citenamefont{Boyd, Gupta, Karsch,
  and Laermann}}]{Boyd:1994np}
\bibinfo{author}{\bibfnamefont{G.}~\bibnamefont{Boyd}},
  \bibinfo{author}{\bibfnamefont{S.}~\bibnamefont{Gupta}},
  \bibinfo{author}{\bibfnamefont{F.}~\bibnamefont{Karsch}}, \bibnamefont{and}
  \bibinfo{author}{\bibfnamefont{E.}~\bibnamefont{Laermann}},
  \bibinfo{journal}{Z. Phys.} \textbf{\bibinfo{volume}{C64}},
  \bibinfo{pages}{331} (\bibinfo{year}{1994}), \eprint{hep-lat/9405006}.

\bibitem[{\citenamefont{Chatterjee et~al.}(2013)\citenamefont{Chatterjee,
  Godbole, and Gupta}}]{Chatterjee:2013yga}
\bibinfo{author}{\bibfnamefont{S.}~\bibnamefont{Chatterjee}},
  \bibinfo{author}{\bibfnamefont{R.}~\bibnamefont{Godbole}}, \bibnamefont{and}
  \bibinfo{author}{\bibfnamefont{S.}~\bibnamefont{Gupta}},
  \bibinfo{journal}{Phys.Lett.} \textbf{\bibinfo{volume}{B727}},
  \bibinfo{pages}{554} (\bibinfo{year}{2013}), \eprint{1306.2006}.

\end{thebibliography}

\end{document}